\documentclass[a4paper,fleqn,usenatbib]{mnras}

\usepackage{newtxtext,newtxmath}

\usepackage[T1]{fontenc}
\usepackage{ae,aecompl}

\usepackage{graphicx}	
\usepackage{amsmath}	
\usepackage{amssymb}	

\usepackage{subfig}

\newcommand*\rad{~rad\,m$^{-2}$}
\newcommand*\sigmaRM{\sigma_{\rm RM}}

\newcommand*\DeltaRM{\Delta{\rm RM}}

\title[Faraday rotation in PKS~J0636$-$2036]{Faraday rotation at low frequencies: magnetoionic material
of the large FRII radio galaxy PKS~J0636$-$2036}

\author[S.~P.~O'Sullivan et al.]{
S.~P.~O'Sullivan,$^{1,2}$\thanks{E-mail: shane@hs.uni-hamburg.de}
E. Lenc,$^{3,4}$
C.~S.~Anderson,$^{5}$
B.~M.~Gaensler,$^{4,6}$
and T.~Murphy$^{3,4}$
\\
$^{1}$Hamburger Sternwarte, Universit\"at Hamburg, Gojenbergsweg 112, Hamburg 21029, Germany\\
$^{2}$Instituto de Astronom\'{i}a, Universidad Nacional Aut\'{o}noma de M\'{e}xico (UNAM), A.P. 70-264, 04510 M\'{e}xico, D.F., M\'{e}xico\\
$^{3}$Sydney Institute for Astronomy, School of Physics, The University of Sydney, NSW 2006, Australia\\
$^{4}$ARC Centre of Excellence for All-sky  Astrophysics (CAASTRO)\\
$^{5}$CSIRO Astronomy \& Space Science, Kensington, Perth 6151, Australia\\
$^{6}$Dunlap Institute for Astronomy and Astrophysics, University of Toronto, 50 St.\ George Street, Toronto, ON M5S 3H4, Canada
}

\date{Accepted XXX. Received YYY; in original form ZZZ}

\pubyear{2017}

\begin{document}
\label{firstpage}
\pagerange{\pageref{firstpage}--\pageref{lastpage}}
\maketitle

\begin{abstract}
We present a low-frequency, broadband polarization study of the FRII radio galaxy PKS~J0636$-$2036 ($z=0.0551$), using the Murchison Widefield Array (MWA) from 70 to 230~MHz. The northern and southern hotspots (separated by $\sim$14.5\arcmin~on the sky) are resolved by the MWA (3\arcmin.3 resolution) and both are detected in linear polarization across the full frequency range. 
A combination of Faraday rotation measure (RM) synthesis and broadband polarization model-fitting are used to constrain the Faraday depolarization properties of the source. 
For the integrated southern hotspot emission, two RM component models are strongly favoured over a single RM component, and the best-fitting model requires Faraday dispersions of approximately 0.7 and 1.2\rad~(with a mean RM of $\sim$50\rad). High resolution imaging at 5\arcsec~with the ATCA shows significant sub-structure in the southern hotspot and highlights some of the limitations in the polarization modelling of the MWA data. 
Based on the observed depolarization, combined with extrapolations of gas density scaling-relations for group environments, we estimate magnetic field strengths in the intergalactic medium between $\sim$0.04 and 0.5\,$\mu$G. 
We also comment on future prospects of detecting more polarized sources at low frequencies. 
\end{abstract}




\section{Introduction}
\label{sec:introduction}

Active galactic nuclei (AGN) can eject jets of relativistic plasma powered by the central supermassive black hole of the host galaxy. These AGN jets have sufficient energy to significantly impact the evolution of the host galaxy and its immediate environment by heating the interstellar and intergalactic gas that might otherwise cool and eventually form stars \citep[e.g.][]{kronberg2001, croton2006}. Determining the exact mechanism(s) by which these AGN jets can transfer their energy to the environment is important for understanding the role of AGN in the evolution of galaxies \citep{fabian2012, mcnamara2012, heckmanbest2014}. Shocks, sound waves, cosmic rays, as well as entrainment and large scale mixing of gas by the jets (and lobes) have been studied as potential mechanisms for depositing the latent AGN jet energy into its environment \citep[e.g.][]{bruggenkaiser2002, fabian2003, ensslin2011, ruszkowski2017}. 

The structure of AGN jets and lobes is illuminated by particle acceleration processes that generate synchrotron emission that is easily detected at radio wavelengths. This synchrotron emission is intrinsically highly linearly polarized, which enables radio spectropolarimetry observations to provide a sensitive probe of thermal plasma via the effect of Faraday rotation. 
The Faraday rotation is detected through the change in the linear polarization angle ($\Delta\psi$) as a function of wavelength-squared ($\lambda^2$), where $\Delta\psi={\rm RM}\lambda^2$, with the Faraday rotation measure (RM) defined as
\begin{equation}
{\rm RM}_{[{\rm rad~m}^{-2}]} = 0.812 \int_{\rm source}^{\rm telescope} n_{e\,\,[{\rm cm}^{-3}]} \,\, B_{||\,\,[{\rm \mu G}]} \,\, dl_{\,\,[\rm{pc}]} 
\label{rmeqn}
\end{equation}
where $B_{||}$ is the line-of-sight magnetic field, $n_e$ is the free electron number density, and $l$ is the path length through the magnetoionic medium.

Since the precision with which the Faraday rotation measure can be determined depends on the total wavelength-squared coverage, low-frequency radio telescopes (with frequencies less than 300~MHz) can outperform traditional centimetre radio facilities by more than two orders of magnitude in this regard \citep{Brentjens:2005}. 
Finding extragalactic polarized sources at low radio frequencies has been challenging to date due to the often poor angular resolution, the influence of the ionosphere, and the expected strong effect of Faraday depolarization \citep[c.f.][]{farnsworth2011, debruyn2012, giebubel2013}. However, recent advancements in low frequency radio interferometers, for example the Muchison Widefield Array \citep[MWA;][]{Tingay:2013} and the Low Frequency Array \citep[LOFAR;][]{vanhaarlem2013} show that it is possible to detect extragalactic polarized sources at these frequencies \citep[e.g.][]{Bernardi:2013, Mulcahy:2014, Jelic:2015, orru2015, Lenc:2016}. 

In this paper, we study the brightest polarized source detected to date at low frequencies, PKS~J0636$-$2036, 
using the MWA telescope from 70 to 230 MHz. This is the first broadband polarization and Faraday rotation analysis of a radio galaxy at such low radio frequencies, providing key constraints on the thermal gas properties of radio galaxies, and important insights for studies in this area with current and future radio telescopes (e.g.~SKA1-low\footnote{http://skatelescope.org/}). 

PKS~J0636$-$2036 is a nearby FRII radio galaxy identified with an isolated elliptical host galaxy at $z=0.0551$ \citep{schilizzimcadam1975, Kronberg:1986}. The radio galaxy has two bright, roughly co-linear hotspots at either ends of the radio structure connected by fainter bridges of radio emission (the lobes), while the host galaxy has broad optical emission lines and extended emission-line gas within 20\arcsec~of the nucleus \citep{Baum:1988}. 
The radio luminosity at 1.4~GHz is $L_{\rm 1.4\,GHz}\sim6\times10^{25}$~W~Hz$^{-1}$. The southern hotspot is the brightest feature of the source, and the southern lobe is $\sim$100\arcsec~longer on the sky than the northern lobe. 
The total size of the radio galaxy is $\sim$870\arcsec, which led it to be listed as a Giant Radio Galaxy (GRG) in early studies \citep{Danziger:1978}, although with the cosmology used in this paper the projected linear size is $\sim$957~kpc. 
GRGs are typically defined as radio sources with linear sizes of $\gtrsim1$~Mpc, 
likely due to their jets propagating in a low density environment \citep[e.g.][]{mack1998, schoenmakers2000, subrahmanyan2008, machalski2011}. 

In Section~\ref{sec:mwaobs}, we describe the MWA observations and data reduction, as well as a description of high angular resolution 
observations with the Australia Telescope Compact Array (ATCA). 
Section~\ref{polmodel} describes our general approach to the broadband polarization modelling. 
Section~\ref{sec:results} presents the results for the northern and southern hotspot of PKS~J0636$-$2036. 
In Section~\ref{interpret} we present various scenarios to explain the broadband polarization data, and 
discuss the likelihoods and implications of these scenarios in Section~\ref{sec:discussion}. 
The conclusions are listed in Section~\ref{sec:conclusion}. 
Throughout this paper, we assume a flat $\Lambda$CDM cosmology with 
H$_0 = 67.3$ km s$^{-1}$ Mpc$^{-1}$, 
$\Omega_M=0.315$ and $\Omega_{\Lambda}=0.685$ \citep{planck2014}.
At the redshift of the source, 1\arcsec~corresponds to a linear size of 1.1~kpc. 
We define the total intensity spectral index, $\alpha$, 
such that the observed total intensity ($I$) at frequency $\nu$ follows the relation 
$I_{\nu}\propto\nu^{\rm{+}\alpha}$.

\begin{table*}
\centering
\caption{Summary of observing specifications for each observing band. The lowest and highest observing frequency are specified by $\nu_{\text{min}}$ and $\nu_{\text{max}}$, respectively. BW$_{\text{chan}}$ specifies the channel bandwidth used for imaging and $t_{\text{int}}$ the overall on-source integration time (note that the ATCA data is spread over two pointings in order to image both the northern and southern hotspot). $\delta\phi$ and $\lvert\phi_{max}\rvert$ are the resolution and Faraday depth range available in each band when using the RM synthesis technique.}
\label{table:obspar}
\begin{tabular}{l c c c c c c c}
\hline\hline
Array & Band & $\nu_{\text{min}}$ & $\nu_{\text{max}}$ & BW$_{\text{chan}}$ & $t_{\text{int}}$ & $\delta\phi$ & $\lvert\phi_{\text{max}}\rvert$ \\
      & (MHz) & (MHz)  & (MHz)   & (MHz) & (min) & (rad\,m$^{-2}$) & (rad\,m$^{-2}$) \\ [0.5ex]
\hline
MWA   & 89   & 74.615  & 102.935 & 0.04  & 12    & 0.45 & 100 \\
MWA   & 118  & 103.135 & 133.655 & 0.04  & 14    & 1.0  & 264 \\
MWA   & 154  & 138.995 & 169.475 & 0.04  & 16    & 2.3  & 647 \\
MWA   & 185  & 169.715 & 200.195 & 0.04  & 16    & 3.9  & 1178 \\
MWA   & 216  & 200.435 & 230.915 & 0.04  & 8     & 6.3  & 1940 \\
ATCA  & 2150 & 1284    & 3012    & 32.0 & 230   & 78   & 661 \\ 
\hline
\end{tabular}
\end{table*}

\begin{table*}
\centering
\caption{Summary of imaging parameters for each observing band. $r$ is the robustness weighting used for imaging. $\sigma$ is the image rms noise for each of the Stokes parameters when utilizing the entire available bandwidth. $\theta_{\text{maj}}$ and $\theta_{\text{min}}$ are the major and minor axis of the synthesized beam (FWHM), respectively. PA is the position angle of the synthesized beam measured from north to east.}
\label{table:imgpar}
\begin{tabular}{l r r r r r r r r r}
\hline\hline
Array & Band & $r$ & $\sigma_{i}$ & $\sigma_{q}$ & $\sigma_{u}$ & $\sigma_{v}$ & $\theta_{\text{maj}}$ & $\theta_{\text{min}}$ & PA \\
      & (MHz) & & (mJy\,beam$^{-1}$)  & (mJy\,beam$^{-1}$)   & (mJy\,beam$^{-1}$) & (mJy\,beam$^{-1}$) &             &             & (deg) \\ [0.5ex]
\hline
MWA   & 89   & -1 & 340 & 35  & 36  & 34  & 4.3\arcmin  & 4.0\arcmin  & $-70$ \\
MWA   & 118  & -1 & 192 & 12  & 12  & 12  & 3.7\arcmin  & 3.5\arcmin  & $-47$ \\
MWA   & 154  & -1 & 159 & 7.6 & 7.8 & 6.5 & 3.5\arcmin  & 3.3\arcmin  & $-74$ \\
MWA   & 185  & -1 & 152 & 6.0 & 6.1 & 5.2 & 3.4\arcmin  & 3.3\arcmin  & $-87$ \\
MWA   & 216  & -1 & 151 & 7.1 & 6.0 & 5.4 & 3.4\arcmin  & 3.3\arcmin  & $-80$ \\
ATCA  & 2150 & 0  & 3.8 & 0.5 & 0.9 & 0.2 & 5.0\arcsec  & 5.0\arcsec  & $0$   \\
ATCA  & 2150 & 0  & 9.2 & 0.8 & 1.0 & 0.3 & 25.0\arcsec & 25.0\arcsec & $0$   \\ 
\hline
\end{tabular}
\end{table*}

\section{Observations and data reduction}
\label{sec:mwaobs}

\subsection{MWA Observations}

The Murchison Widefield Array (MWA) is located at the Murchison Radio Observatory in Western Australia. At the time of observation, the MWA consisted of 128 tiles (with a maximum baseline length of $\sim3$ km), where each tile has a regular $4\times4$ grid of dual-polarization dipoles \citep{Tingay:2013}.

We used archival\footnote{\url{http://mwa-metadata01.pawsey.org.au/admin/observation/observationsetting/}} GLEAM (A Galactic and Extragalactic All-Sky MWA) survey visibility data ranging from 72 MHz to 231 MHz \citep{Wayth:2015,Hurley-Walker:2017}. Our primary target, PKS~J0636$-$2036, drifted within the half-power beam-width on 2013 November 25 between 17:37 and 19:13 UTC. 
The observing specifications are listed in Table~\ref{table:obspar}. 

The data was flagged using \textsc{aoflagger} \citep{Offringa:2012} to remove radio frequency interference (RFI) and channels affected by the poly-phase filter bank \citep{Ord:2015}. In all, approximately 25\% of data was flagged although only a small fraction of this was due to actual RFI \citep{Offringa:2015}.

We used the real-time calibration and imaging system \citep[\textsc{rts};][]{Mitchell:2008, Ord:2010} for bandpass and gain calibration on a per 2-minute snapshot basis. We used a local sky model centered on the inner $20\degr$ of the primary beam and taking an ensemble of the 30 most significant sources taken from the MWA Commissioning Survey \citep{Hurley-Walker:2014}. Visibility data from baselines shorter than $50\lambda$ were down-weighted to improve calibration in the presence of diffuse structure. We estimate the uncertainty on the absolute calibration to be better than 10\%.

Imaging was also carried out with the \textsc{rts} with a taper to down-weight baselines above 700$\lambda$ and a lower baseline length limit of 50$\lambda$. The baseline restrictions provide a near-constant beam size across all frequency bands (Table \ref{table:imgpar}) and limit  contamination from large scale structure. To avoid bandwidth depolarization, all bands were imaged at the full 40 kHz spectral resolution. 

To correct for ionospheric Faraday rotation we estimated the ionospheric contribution to the RM with \textsc{RMextract}\footnote{\url{https://github.com/maaijke/RMextract}}. The ionospheric component of the RM was then de-rotated from the Stokes $Q$ and $U$ image cubes in each 2 minute snapshot. Once corrected, the image cubes were integrated in time.
Predictions of ionospheric Faraday rotation are typically accurate to within $\sim$$0.2$\,rad\,m$^{-2}$ \citep{Lenc:2016}. For the night-time, near-zenith observations presented here, \citet{lenc2017} showed that the predictive accuracy of the variable component of ionospheric RM was $\sim$$0.03$\,rad\,m$^{-2}$. Similarly, they demonstrated that if the accuracy were worse than $\sim$$0.085$\,rad\,m$^{-2}$ then the southern hotspot would completely depolarize in the 89\,MHz MWA band. These tests do not preclude the possibility of a constant offset error in the ionosphere RM prediction, however, as all MWA bands were observed over the same period they would all exhibit the same offset. 

\begin{table*}
\centering
\caption{Summary of polarization properties of the southern (subscript s) and northern (subscript n) hotspots in each of the observed bands. $S$ is the total intensity, $P$ is the polarized intensity, $p$ is the polarized fraction and RM is the  measured Faraday rotation.}
\label{table:rmsummary}
\begin{tabular}{l c c c c c c c c c}
\hline\hline
Array & Band & $S_{\text{s}}$ & $P_{\text{s}}$ & $p_{\text{s}}$ & RM$_{\text{s}}$ & $S_{\text{n}}$ & $P_{\text{n}}$ & $p_{\text{n}}$ & RM$_{\text{n}}$  \\
      & (MHz) & (Jy) & (mJy) & & (rad\,m$^{-2}$)  & (Jy) & (mJy) & & (rad\,m$^{-2}$) \\ [0.5ex]
\hline
MWA   & 89   & 29.6 & 262  & 0.9\%  & $+49.4\pm1.2$  & 14.9 & 169 & 1.1\% & $+35.9\pm0.03$ \\
MWA   & 118  & 22.6 & 451  & 2.0\%  & $+50.0\pm0.02$ & 10.2 & 146 & 1.4\% & $+36.0\pm0.03$ \\
MWA   & 154  & 17.8 & 558  & 3.1\%  & $+50.1\pm0.04$ & 7.8  & 215 & 2.7\% & $+36.3\pm0.03$ \\
MWA   & 185  & 15.7 & 985  & 6.3\%  & $+50.1\pm0.03$ & 6.8  & 192 & 2.9\% & $+36.5\pm0.1$ \\
MWA   & 216  & 13.9 & 1283 & 9.2\%  & $+50.2\pm0.05$ & 6.0  & 142 & 2.4\% & $+36.1\pm0.3$ \\
NVSS  & 1400 & 3.01 & 456  & 15.1\% & $+47.1\pm1.9$  & 1.53 & 164  & 10.7\% & $+34.8\pm7.2$ \\
\hline
\end{tabular}\\
\scriptsize{Note: NVSS RM values not corrected for the ionospheric RM contribution.}
\end{table*}

\begin{figure*}
\begin{center}
\includegraphics[width=0.45\linewidth]{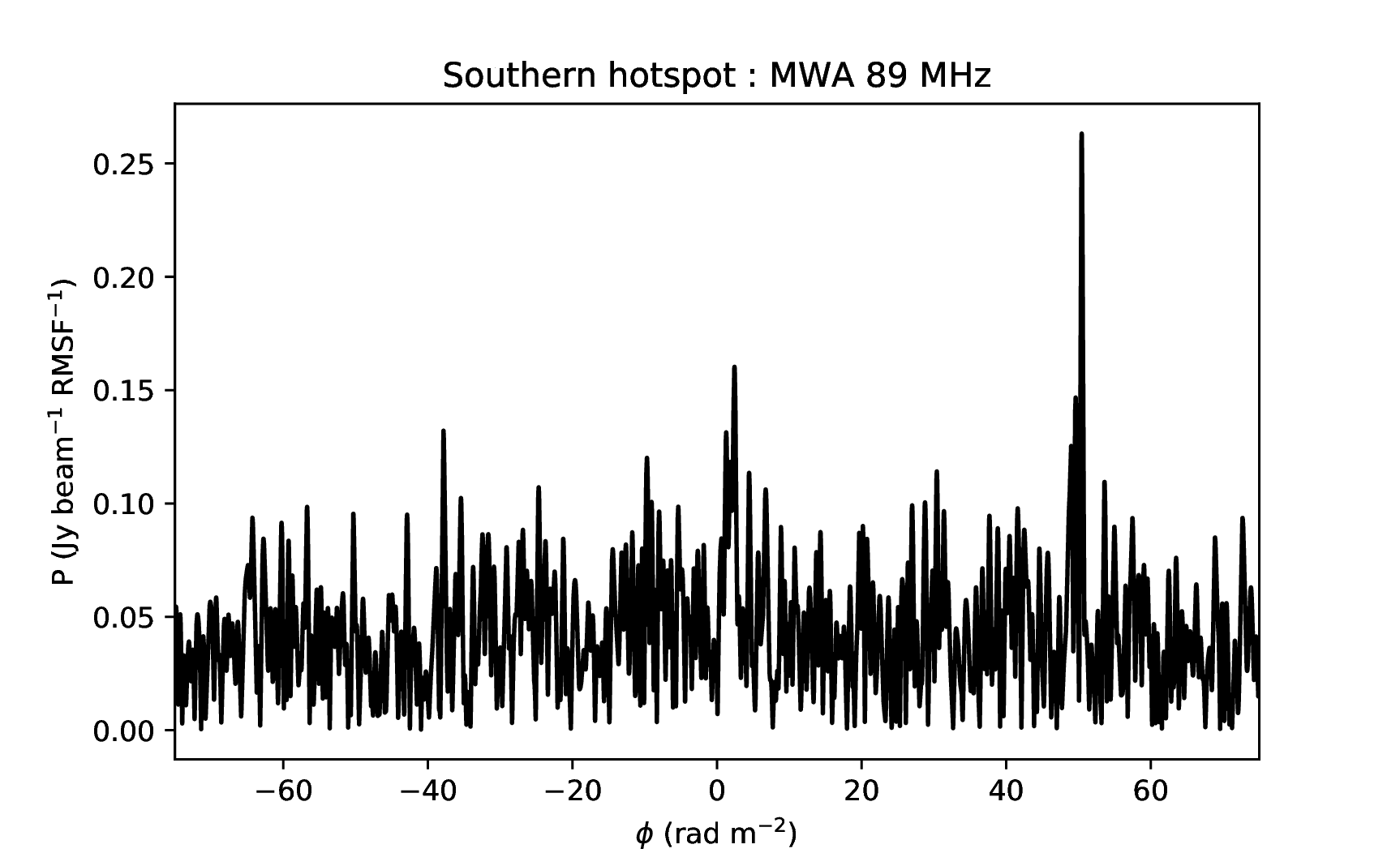}
\includegraphics[width=0.45\linewidth]{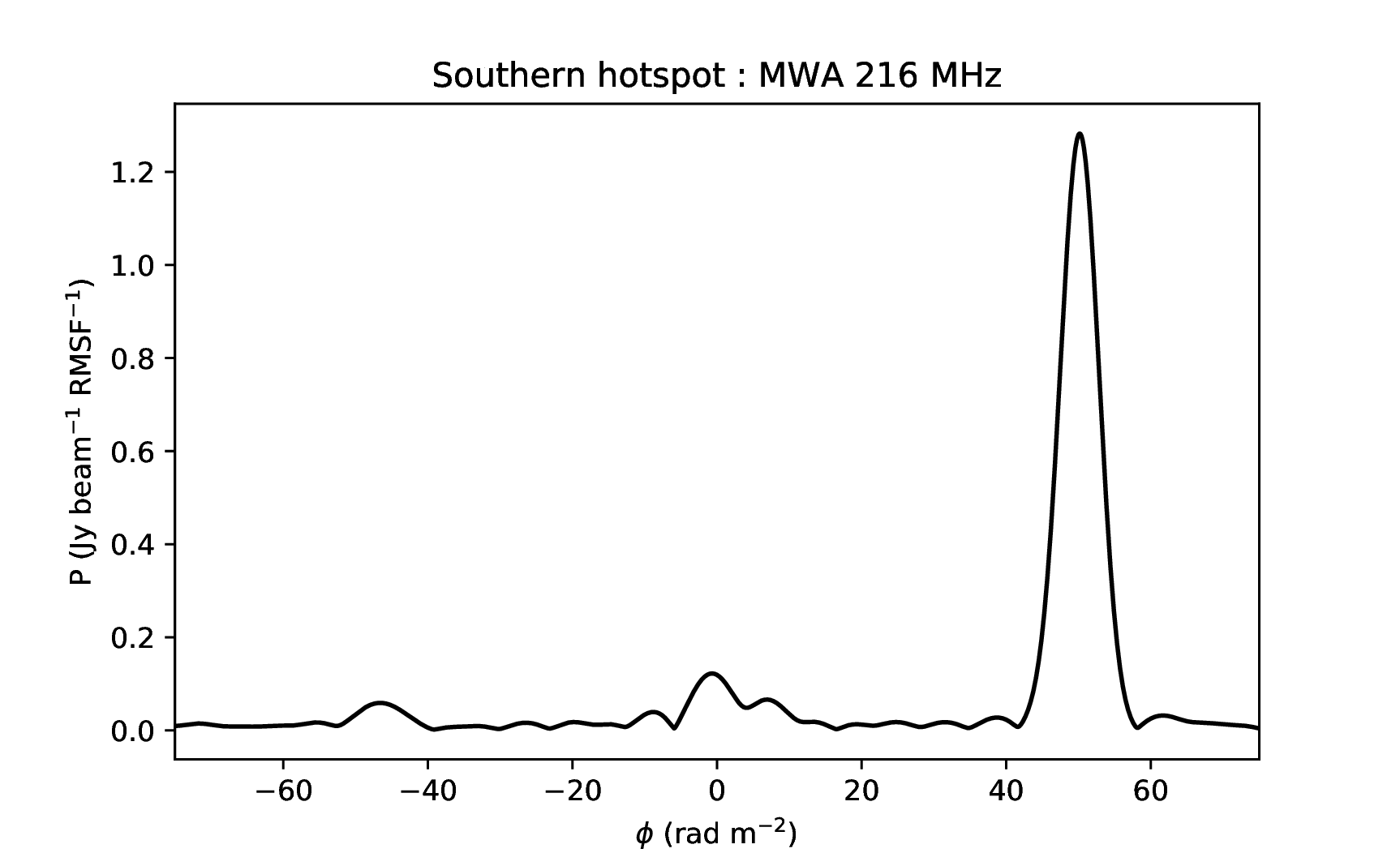}
\caption{Faraday dispersion functions for the southern hotspot of PKS~J0636$-$2036, shown for the lowest (left) and highest (right) frequency GLEAM bands.}
\label{fig:fdffreq}
\end{center}
\end{figure*}

\subsection{ATCA Observations}
\label{sec:atcaobs}
PKS J0636-2036 was observed using Directors Time (project CX317) on 2015 February 27 from 04:40-14:47 UTC using the Australia Telescope Compact Array (ATCA) in a 6 km configuration (configuration 6C). The Compact Array Broadband Backend \citep[CABB;][]{Wilson:2011} was used with 2048 MHz bandwidth (1 MHz channels) centered at 2.1 GHz (16 cm). In total, $12 \times 10$ minute scans were taken of the southern hotspot and $11\times 10$ minutes scans of northern hotspot. 

A 15 minute scan of the standard flux density calibrator source PKS B1934-638 was used for calibration. Primary flux density calibration and data flagging for radio frequency interference was performed using standard calibration procedures for the ATCA in the data reduction package Miriad \citep{Sault:1995}. The observations suffered from significant RFI at the lower end of the band and the first $\sim$$150$ channels were flagged.

Time-dependent calibration was performed by first generating spectral models for both the northern and southern hotspots with the task \textsc{mfclean} and then using this model to perform phase-only self-calibration. Once calibrated, the visibilities were averaged down to 64 $\times$ 32\,MHz channels and full-Stokes image cubes were generated and deconvolved using the task \textsc{clean}. The resulting image cubes were restored with a $5\arcsec$ beam to enable the high resolution structure of the polarized hot spot to be studied. They were also restored with a $25\arcsec$ beam to study the unresolved properties of the hotspot.

To determine the effect of ionospheric Faraday rotation we used \textsc{RMextract} to estimate the RM component of the ionosphere over the course of the ATCA observing period. The mean shift in RM as a result of ionospheric Faraday rotation was estimated to be $-2.5$\,rad\,m$^{-2}$ and all RM measurements made with the ATCA were adjusted to account for this.

\section{polarization model-fitting approach}
\label{polmodel}
In order to model the broadband polarization data, 
we consider several different polarization models that describe different types of Faraday depolarization. In particular, we follow the algorithm described in \cite{osullivan2017}, where they use a general model that can describe Faraday effects internal or external to the emission region, from both uniform and turbulent magnetic fields. The complex polarization equation for this model is 
\begin{equation}
P = p_{0} \, e^{2 i (\psi_{0}+{\rm RM} \lambda^2)} \frac{\sin \Delta {\rm RM} \lambda^2}{\Delta {\rm RM} \lambda^2} e^{-2\sigma^2_{{\rm RM,B}} \lambda^4}
\label{burneqn}
\end{equation}
where for each complex polarization component ($P$) or `RM component', $p_{0}$ is the intrinsic degree of polarization and $\psi_{0}$ is the intrinsic polarization angle.
The parameter $\Delta{\rm RM}$ usually describes internal Faraday depolarization in the presence of a uniform magnetic field, but can also be consistent with external Faraday depolarization caused by a linear gradient in RM across the emission region \citep[e.g.][]{sokoloff1998, schnitzeler2015}.  
The parameter $\sigma_{\rm RM,B}$ is used to describe Faraday depolarization from RM variations due to a turbulent magnetic field, a process often referred to as external Faraday dispersion or `Burn-law' depolarization \citep{Burn:1966, laing2008}. 

As described in \cite{Tribble:1991} and \cite{sokoloff1998}, the assumptions underlying the `Burn-law' are expected to break down in the long wavelength regime ($2\sigma_{\rm RM}\lambda^2 \gg 1$, $p(\lambda^2)/p(0) \ll 0.5$). Therefore, we also consider polarization models that include the expected transition from an exponential decline to a power-law decline in $p(\lambda^2)$, as described in eqn.~3 of \cite{tribble1992}. In this case, 
\begin{equation}
P = \mathcal{P} \frac{\sin \Delta {\rm RM} \lambda^2}{\Delta {\rm RM} \lambda^2} \left( \frac{ 1-e^{-S-4\sigma_{\rm RM,T}^2\lambda^4} }{ 1+4\sigma_{\rm RM,T}^2\lambda^4/S } + e^{-S-4\sigma_{\rm RM,T}^2\lambda^4}\right)^{1/2}
\label{tribbleeqn}
\end{equation}
where $\mathcal{P}=p_{0} \, e^{2 i (\psi_{0}+{\rm RM} \lambda^2)}$ and $\sigma_{\rm RM,T}$ is used to describe Faraday depolarization following the `Tribble-law', with the parameter $S=s_0/t$ depending on the resolution of the observations ($t$) and the scale size of the RM fluctuations ($s_0$). For example, as $S\rightarrow0$ the RM structure becomes fully unresolved and the `Burn-law' behaviour is recovered, while as $S\rightarrow \infty$ the RM structure becomes fully resolved and there is no depolarization. 

In the `short-wavelength' regime ($p(\lambda^2)/p_0 > 0.5$), the $\sigmaRM$ and $\DeltaRM$ parameters produce equivalent amounts of depolarization for $\DeltaRM \sim 3.22\sigmaRM$. Furthermore, the $\DeltaRM$ parameter can describe a linear gradient in RM across a flat beam profile, but to describe the more realistic case of a Gaussian beam profile $\DeltaRM$ needs to be divided by a factor of 1.35 \citep{sokoloff1998}.
In the case of internal Faraday rotation, this correction factor is not required. 

In this work, we consider up to a maximum of two RM component models, using Eqns.~2 \& 3 and combinations thereof (e.g.~$P=P_1+P_2$).
To evaluate the quality of fit and to discriminate between models with different numbers of parameters, we use the reduced-$\chi$-squared values ($\chi^2_r$) and the Bayesian Information Criterion (BIC), as described in \cite{osullivan2012}. The model with the lowest BIC and $\chi^2_r$ values was selected as the best fitting model \citep[e.g.][]{raftery1995}. 

\begin{figure*}
\includegraphics[width=\linewidth]{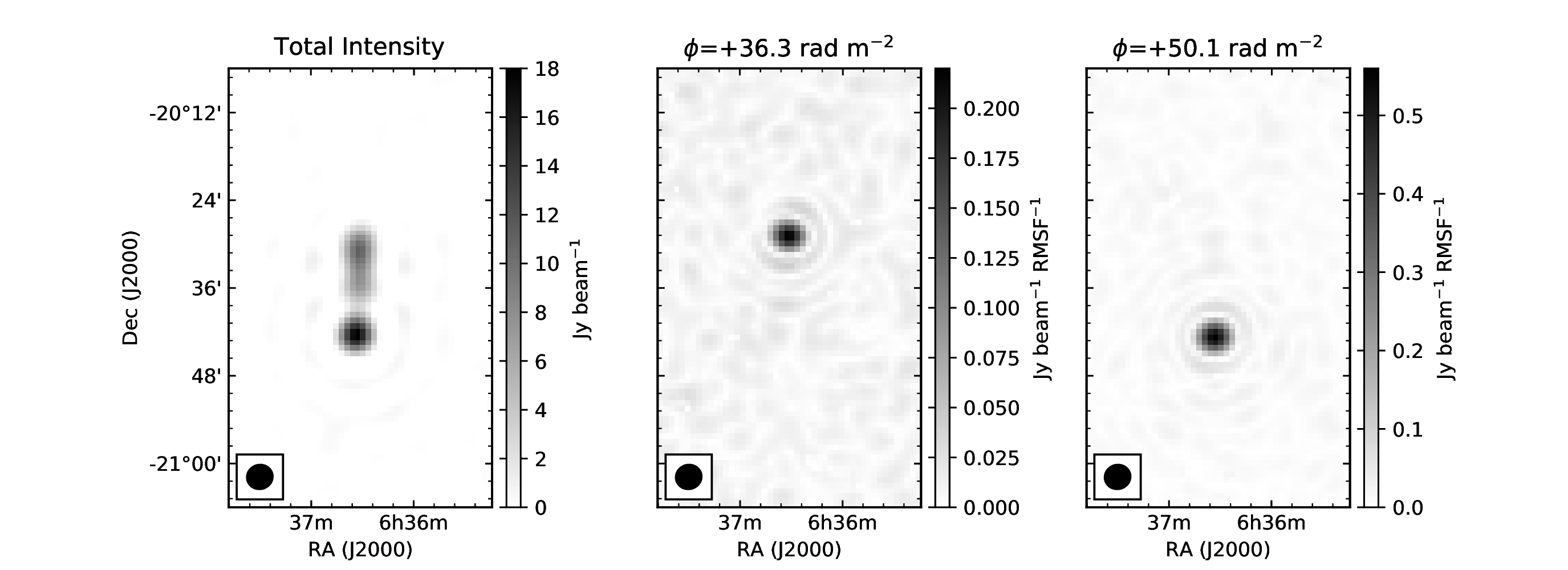}
\caption{Total intensity map of PKS~J0636$-$2036 and the associated polarized intensity maps of the source taken at Faraday depths of $\phi$=+36.3\,rad\,m$^{-2}$ and +50.1\,rad\,m$^{-2}$ (corrected for ionospheric effects). All images were processed using robust visibility weighting (robustness$=-1$) in the $154$\,MHz band.}
\label{fig:polim}
\end{figure*}

\section{Results}
\label{sec:results}

Rotation Measure (RM) Synthesis \citep{Brentjens:2005} was used to determine the Faraday dispersion function (FDF), which is the polarized intensity as a function of Faraday depth \citep{Burn:1966}, for each pixel in each image cube. To decrease the effect of sidelobes in Faraday space, the RM cubes were deconvolved using RM CLEAN \citep{Heald:2009}. With the combined ionospheric calibration and RM synthesis technique, polarization was detected in both the northern and southern hot spot of PKS~J0636$-$2036 across each of the five MWA observing bands and in the ATCA data 
(faint polarized emission was also detected in the northern lobe/bridge in the two highest MWA bands, and is summarised in Section~\ref{nh}). 

The peak polarized intensity and RM was measured at J2000 06$^{\rmn{h}}$36$^{\rmn{m}}$33$\fs$15, $-20$\degr 42$\arcmin$40$\arcsec$ for the southern hotspot and 06$^{\rmn{h}}$36$^{\rmn{m}}$31$\fs$75, $-20$\degr 29$\arcmin$00$\arcsec$ for the northern hotspot in each of the RM cubes and is summarised in Table \ref{table:rmsummary}. The Faraday dispersion function for the southern hotspot of PKS~J0636$-$2036, which is the brighter of the two hotspots, is shown in Figure \ref{fig:fdffreq} for the lowest and highest MWA frequency bands.
Figure \ref{fig:polim} shows the total intensity image of PKS~J0636$-$2036 and the polarized intensity maps for the northern and southern hot spots in the 154\,MHz band at $+36.3$ and $+50.1$\,rad\,m$^{-2}$, respectively. 

\begin{figure}
\begin{center}
\includegraphics[width=0.9\columnwidth]{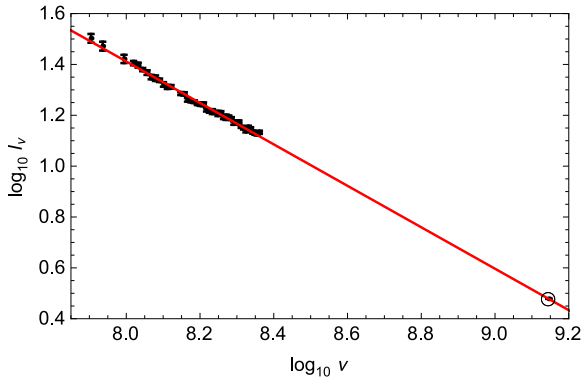}
\caption{Stokes $I$ flux versus frequency for the southern hotspot of PKS~J0636$-$2036, from MWA and NVSS at 3\arcmin.3 resolution. 
The NVSS data point is highlighted by a circle in the bottom left corner with an error of 0.5~mJy~beam$^{-1}$. 
The best fitting spectral index value is $\alpha=-0.815\pm0.007$.%
}
\label{SHspix}
\end{center}
\end{figure}

\begin{figure*}
\includegraphics[clip=true, trim=0cm 0cm 0cm 0cm, width=1.0\textwidth]{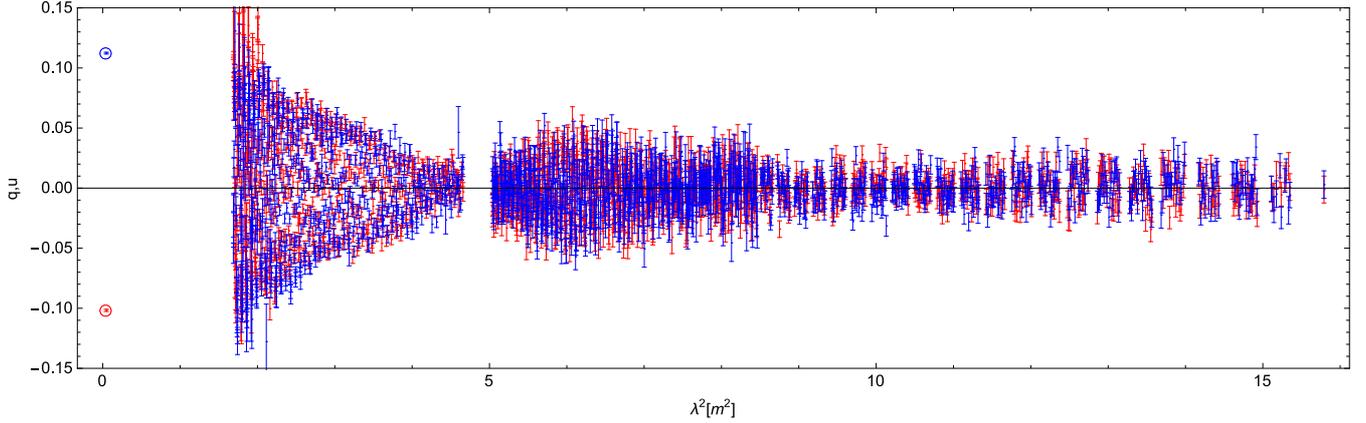}
\caption{Variation in $q$ (red) and $u$ (blue) as a function of wavelength-squared ($\lambda^2$), for the southern hotspot. All data points are the raw 40~kHz MWA channel data from 75 to 230 MHz, except for the left-most points on the plot which are the NVSS $q$, $u$ data at 1.4~GHz (highlighted by the red and blue circles). }
\label{fig:qu}
\end{figure*}

\subsection{Southern Hotspot}
\label{sh}
Figure~\ref{SHspix} shows the total intensity as a function of frequency for the MWA data of the southern hotspot, and also includes the NRAO VLA Sky Survey (NVSS) data at 1.4~GHz \citep{condon1998}. 
The NVSS $I$, $Q$ and $U$ images at 1.4 GHz\footnote{http://www.cv.nrao.edu/nvss/postage.shtml} were convolved to the same beamsize as the MWA data ($\sim$3\arcmin.3). The NVSS total intensity and fractional polarization data were then extracted at the same location as the peak polarized intensity of the southern hotspot in the MWA data (Table~\ref{table:rmsummary}). This also provided a crucial additional data point at $\lambda^2 \simeq 0.046$~m$^2$ to help constrain the polarization model-fitting. 
The solid line in Figure~\ref{SHspix} shows the linear regression fit to the MWA and NVSS total intensity data, with a best-fitting total intensity spectral index $\alpha=-0.815\pm0.007$. There is no evidence for a turnover/cutoff in the total intensity spectrum down to 75~MHz; although we cannot rule out higher resolution observations at low frequencies revealing more interesting spectral structure for the hotspot \citep[e.g.][]{harwood2016, harwood2017}. 

The total RM of the southern hotspot of $\sim$50\rad~(Table~\ref{table:rmsummary}) is dominated by the Galactic foreground \citep{Kronberg:1986, taylor2009}, with the RM contribution local to the source expected to be much smaller given how far the source extends outside its host galaxy, in addition to the low galaxy-density environment in which the source resides \citep{Baum:1988}. 
Thus, one of the best ways to study the magnetoionic material local to the source is through the Faraday depolarization observed across the MWA bands. 

Figure~\ref{fig:qu} shows the MWA broadband polarization data in each channel for the southern hotspot (with the NVSS data point included). The $q$, $u$ data oscillate rapidly because of the large RM, however, from the envelope of the data it is clear that there is significant depolarization across the MWA band, which ranges in wavelength-squared from $\sim$1.7 to 16~m$^2$. For comparison, this is $>300$ times the wavelength-squared coverage in typical cm-wavelength observations with the ATCA \citep[e.g.][]{osullivan2012}. 

Due to the lack of an absolute polarization angle calibrator, there are unknown offsets in the polarization angles measured between the MWA bands. This precludes us from doing $qu$-fitting on the full dataset. We expect that this can be rectified in future when more brightly polarized sources are found and studied at low frequencies. Instead, we model the $p(\lambda^2)$ data (see Section~\ref{sec:depolmodel}) in conjunction with the RM synthesis results within each band (which are robust against the absolute polarization angle calibration). 

\begin{figure}
\begin{center}
\includegraphics[width=0.95\columnwidth]{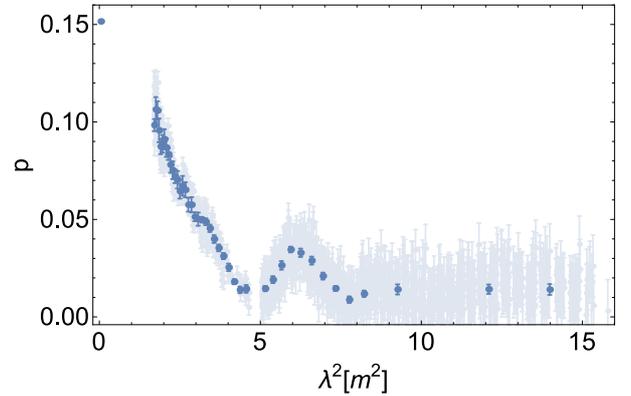}
\caption{Plot of $p(\lambda^2)$ for the RM synthesis averaged data (blue points), after correcting for both the instrumental polarization and polarization bias. 
The transparent points show the raw MWA 40\,kHz channel data. 
The NVSS data is included at $\lambda^2 \simeq 0.046$~m$^2$.
}
\label{pwithleakage}
\end{center}
\end{figure}

To construct a reliable $p(\lambda^2)$ depolarization curve, we need to correct for both the instrumental polarization and the polarization bias. In order to isolate most of the instrumental polarization (which shows up mainly at ${\rm RM}\sim0$\rad) from the real signal (at ${\rm RM}\sim50$\rad), we used RM synthesis to coherently average the polarization data in small intervals across each MWA band (with 10 intervals across each band). We only retained those data points with signal to noise greater than 6 \citep{Brentjens:2005}, and also corrected the degree of polarization for the effect of polarization bias, following \cite{george2012}. 
By extracting the fractional polarization values at the peak in the RM synthesis spectrum near $\sim50$\rad~, we were able to obtain a cleaner signal than if we just took the raw channel fractional polarization values. 
Figure~\ref{pwithleakage} shows the resulting $p(\lambda^2)$ from the coherently averaged data from RM synthesis (blue points), overlaid on the raw channel data (uncorrected for bias and the effect of instrumental polarization). 

\subsubsection{ATCA imaging}
\label{atcaresults}
\begin{figure}
\begin{center}
\includegraphics[width=0.85\columnwidth]{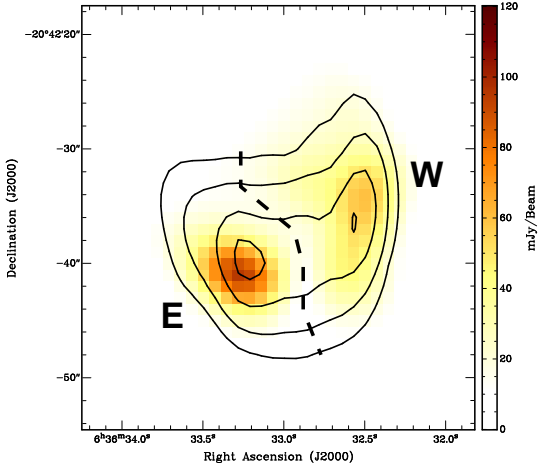}
\caption{Total intensity (contours) and polarized intensity (colour-scale) of southern hotspot of PKS~J0636$-$2036 at 5\arcsec~resolution with ATCA at 16~cm band. 
The dashed line denotes the separation between the eastern region (Region E) and the western region (Region W) of the hotspot.
}
\label{Ip}
\end{center}
\end{figure}

The ATCA data at 16~cm enabled high angular resolution imaging of the southern hotspot at 5\arcsec. 
This allowed us to resolve the total intensity and polarization structure, shown in Figure~\ref{Ip}. The size of the hotspot region is $\sim$20~kpc, which is consistent with expectations based on self-similar models of FRII radio galaxies of this size \citep[e.g.][]{hardcastle1998b, kaiseralexander1997}. 
While the current ATCA data is excellent for resolving the structure of the hotspot, it is not sufficient for recovering all of the more diffuse lobe emission. Indeed the integrated total intensity emission is $\sim$15\% lower than expected from extrapolation of the best fitting line in Figure~\ref{SHspix}. Therefore, we only use the NVSS data as a more reliable `short-wavelength' constraint in the polarization model fitting (Section~\ref{sec:depolmodel})

\begin{figure}
\begin{center}
\includegraphics[width=0.85\columnwidth]{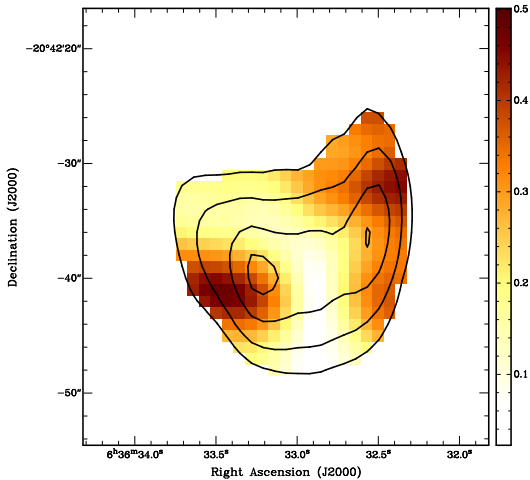}
\caption{Total intensity (contours) and polarized fraction (colour-scale) of southern hotspot of PKS~J0636$-$2036 at 5\arcsec~resolution with ATCA at 16~cm band.%
}
\label{Ifp}
\end{center}
\end{figure}

RM synthesis was used on a pixel-by-pixel basis across the hotspot emission to obtain the distributions of RM (corrected for ionospheric Faraday rotation; see Section~\ref{sec:atcaobs}) and fractional polarization, as well as the polarized intensity with the sensitivity of the full bandwidth. 
Figure~\ref{Ip} shows how the total intensity and polarization structure of the southern hotspot is split into two distinct regions (Region E \& Region W), which is consistent with a previous VLA image at 5~GHz with 4\arcsec.5 resolution \citep{Kronberg:1986}. The peak polarized intensities of the two regions are slightly offset from the peak in total intensity, and the degree of polarization (Figure~\ref{Ifp}) increases sharply towards the outer edges of the regions, reaching maximum values of $\sim45\%$. 
This type of structure is not uncommon for FRII hotspots \citep[e.g.][]{rudnick1981, rudnick1988, leahy1997}, where at high angular resolution  they are often resolved into `primary' and `secondary' hotspot regions \citep{laing1989}. 

\begin{figure}
\begin{center}
\includegraphics[width=0.91\columnwidth]{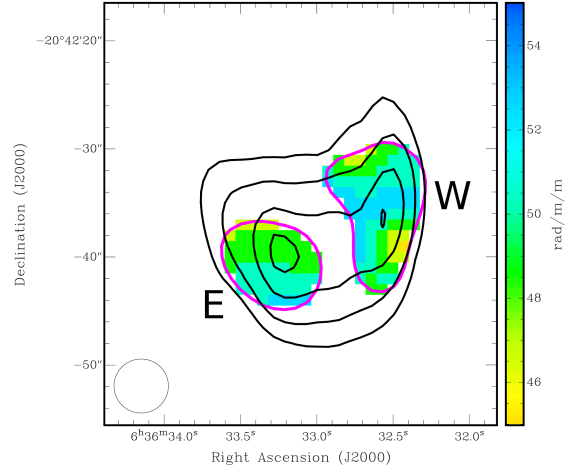}
\caption{Total intensity (contours) and rotation measure distribution (colour-scale) of southern hotspot of PKS~J0636$-$2036 at 5\arcsec~resolution with ATCA at 16~cm band. The magenta contours outline where the polarized intensity is greater than 10$\sigma_{QU}$ for the two regions (E \& W). 
}
\label{IpRM}
\end{center}
\end{figure}

\begin{figure}
\begin{center}
\includegraphics[width=0.8\columnwidth]{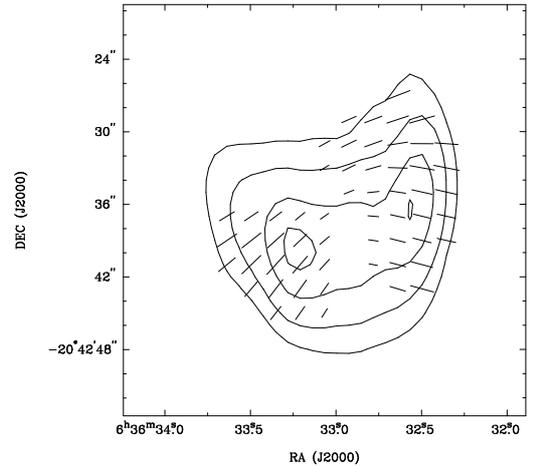}
\caption{Total intensity (contours) and intrinsic polarization angle (sticks) of southern hotspot of PKS~J0636$-$2036 at 5\arcsec~resolution with ATCA at 16~cm band.%
}
\label{IX0}
\end{center}
\end{figure}

The RM distribution is shown in Figure~\ref{IpRM}, with the RM only shown where the polarized intensity is above $10\sigma_{QU}$ (where $\sigma_{QU}$ is estimated from the rms noise level in the Faraday dispersion function of $Q$ and $U$ in regions far from the peak). 
The RM increases across Region~E from the north-east to the south-west, while in Region~W the RM peaks near the middle and decreases to the north-east and south-west.
The intrinsic polarization angle distribution (Figure~\ref{IX0}) was obtained by de-rotating the polarization angles to zero wavelength using the RM value of each pixel. This shows that the two brightly polarized regions have different intrinsic polarization angle distributions, and that the polarization angles are approximately orthogonal to the total intensity contours along the edges. This is consistent with the often observed tendency for the projected magnetic field direction to wrap around the edges of the hotspot region (see e.g.~3C\,445 in \cite{leahy1997} for a similar overall source structure).


\subsubsection{ATCA constraints for broadband depolarization modelling}
\label{atcadepol}
We find no evidence for depolarization within the 16~cm band, by comparing the fractional polarization found using RM synthesis on the upper ($>2.1$~GHz) and lower ($<2.1$~GHz) frequency ranges of the band. 
Therefore, the goal here is to extract quantitative information from the ATCA data that can be used as constraints for the broadband polarization model-fitting of the MWA data (Section~\ref{sec:depolmodel}). 
 
We consider both the integrated polarization and RM properties of the entire hotspot as well as the integrated properties of the two distinct regions of polarization, as identified in Figure~\ref{Ip}. 
By integrating the Stokes $Q$ and $U$ within these two regions (bounded by the dividing line and outermost Stokes $I$ contour), 
we estimate the degree of polarization ($p_{\rm ATCA}$) of each region with respect to the integrated emission of the entire hotspot (as this is the quantity determined by the model fitting of the MWA data). We find the degree of polarization of the region to the west (Region W) to be $\sim$17\%, and the region to the east (Region E) to be $\sim$15\% (see Table~\ref{table:ATCAdata}). 

In order to inform our polarization modelling of the MWA data, we consider polarization-weighted quantities from the ATCA data. 
We estimate the polarization-weighted mean RM (${\rm RM_{wtd}}$) and polarization-weighted RM dispersion ($d{\rm RM_{wtd}}$), respectively, as 
\begin{equation}
{\rm RM_{wtd}} = \frac{ \sum_j p_j {\rm RM}_j }{ \sum_j p_j },
\label{meanrmeqn}
\end{equation}
and
\begin{equation}
d{\rm RM_{wtd}} = \left( \frac{ \sum_j p_j \left({\rm RM}_j - {\rm RM_{wtd}}\right)^2 }{ \sum_j p_j } \right)^{1/2},
\label{stdevrmeqn}
\end{equation}
where the RM is summed over all pixel locations ($j$) and weighted by the polarized intensity ($p$) at each pixel. 
The polarization-weighted mean and standard deviations of the intrinsic polarization angles ($\psi_{\rm 0,wtd}$ and $d\psi_{\rm 0,wtd}$) were obtained in a similar manner. We only sum over those pixels for which we have obtained RM values (i.e.~above a 5$\sigma_{QU}$ cutoff).

Table~\ref{table:ATCAdata} summarises the parameter values calculated in the above manner for the resolved hotspot regions at 16~cm. 
The variation in RM of $d{\rm RM_{wtd}}=2.2$\rad~is relatively small compared to the mean polarization-weighted RM error of 1.4\rad. However, following \cite{leahy1986}, we find a reduced chi-squared of 3.2 for the significance of the observed RM variations for our $5\sigma_{QU}$ cut-off (3.1 for a $10\sigma_{QU}$ cut-off). A reduced chi-squared of $\sim$1 is expected if noise errors dominate the RM fluctuations. 
By subtracting in quadrature the mean polarization-weighted RM error from $d{\rm RM_{wtd}}$, we get an estimate for the true underlying RM dispersion of 1.7\rad. 
However, in Section~\ref{sec:depolmodel} we find that this is an overestimate of the Faraday dispersion required to explain the observed depolarization for the MWA band. 

The intrinsic polarization orientation displays large variations across the hotspot with $\psi_{\rm 0,wtd}$ differing by $\sim52\degr$ between Regions~E and W. 
Within Region~E, there is a relatively small systematic variation in its intrinsic polarization orientation ($\lesssim20\degr$), while Region~W displays a larger systematic change in the field orientation of $\sim40\degr$ from the south towards the north-eastern edge (Fig.~\ref{IX0}). 

In our approach to modelling the MWA data (Sections~\ref{polmodel} \&~\ref{sec:depolmodel}), we assume that each `RM component' is well characterised by a constant intrinsic polarization angle. 
This assumption of distinct regions of constant intrinsic field orientation dominating the emission is, in detail, inconsistent with the observed high resolution polarization structure. This is not overly surprising given the complex total intensity and polarization structure observed in hotspots at high angular resolution \citep[e.g.][]{leahy1997, tingaylenc2008}. 
Additionally, we expect some diffuse polarized emission has been resolved out in our ATCA observations that is present in the MWA data. 
This highlights some limitations of our polarization modelling approach in the presence of significant underlying polarization and Faraday rotation structure. However, in the context of hotspots which can have significant structure on scales of 10's of parsecs \citep[see][and references therein]{tingaylenc2008}, this will remain an issue even with high angular resolution spectropolarimetry at low frequencies. The development and application of more detailed polarization models is required but beyond the scope of the current work.

\begin{table}
\centering
\caption{Summary of ATCA data for southern hotspot.}
\label{table:ATCAdata}
\begin{tabular}{l c c c c c c}
\hline\hline
                 & $p_{\rm ATCA}$     & RM$_{\rm wtd}$      & $d$RM$_{\rm wtd}$    & $\psi_{\rm 0,wtd}$   & $d\psi_{\rm 0,wtd}$   \\
                 & (\%)                             & (rad\,m$^{-2}$)  & (rad\,m$^{-2}$)   & (deg)                & (deg)                     \\ [0.5ex]
\hline
Region E    & 14.7                           & 49.4                  & 1.8                     & $-43.6$         &         5.7    \\
Region W    & 17.3                           & 50.1                  & 2.3                     & $-96.0$         &        11.5   \\
All              & 15.2                           & 49.7                  & 2.2                     & $-71.1$          &         27.8  \\
\hline
\end{tabular}
\scriptsize{Note: Col.~1: Region of hotspot. Col.~2: Degree of polarization at 16~cm. Col.~3: polarization-weighted RM. 
                  Col.~4: polarization-weighted standard deviation in RM. Col.~5: polarization-weighted intrinsic polarization angle. 
                  Col.~6: polarization-weighted standard deviation in the intrinsic polarization angle.}
\end{table}


\subsubsection{Depolarization modelling}
\label{sec:depolmodel}
We applied our polarization model-fitting techniques (Section~\ref{polmodel}) to the RM synthesis-averaged $p(\lambda^2)$ data (Section~\ref{sh}) in order to investigate the physical cause of the observed depolarization behaviour. We fit the polarization models to the $p(\lambda^2)$ data instead of the $q(\lambda^2)$ and $u(\lambda^2)$ data because we lack an absolute calibration of the polarization angle across the five MWA bands. 

\begin{figure} 
\centering
    \includegraphics[angle=0, clip=true, trim=0cm 0cm 0cm 0cm, width=.45\textwidth]{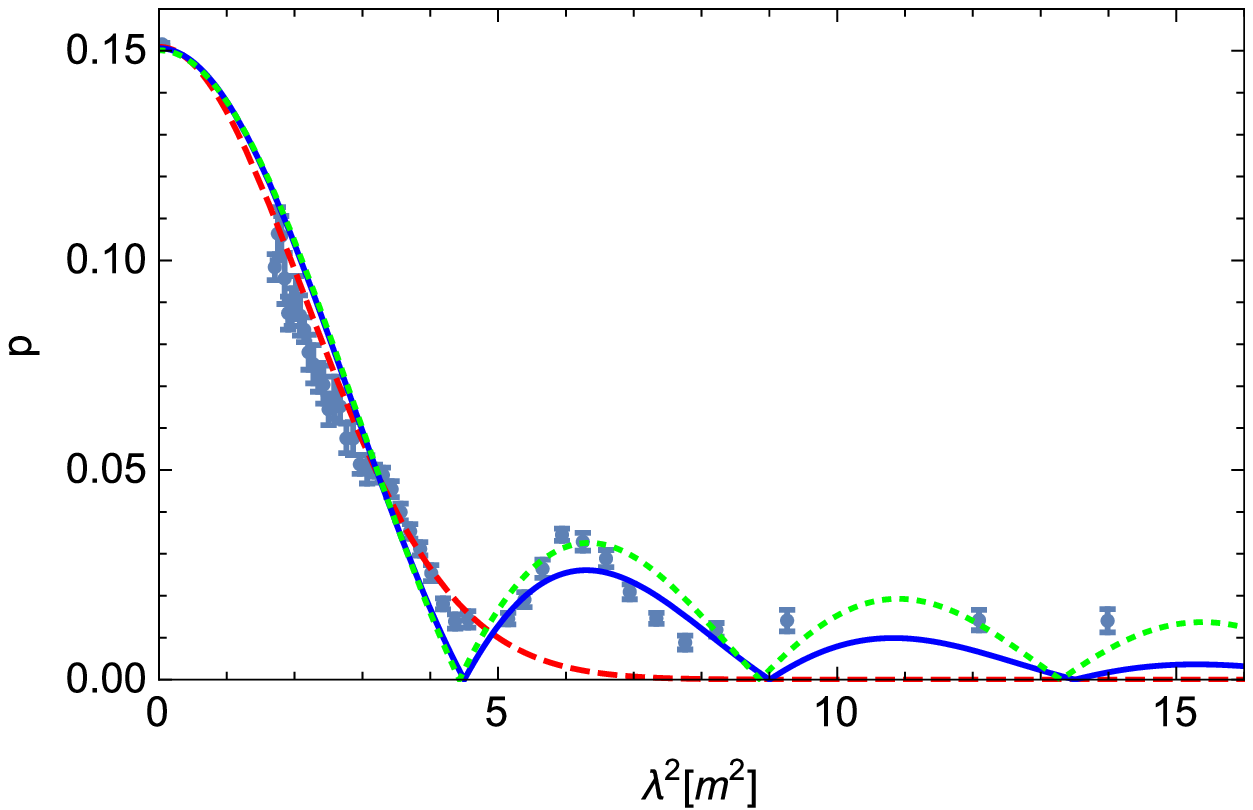}
    \includegraphics[angle=0, clip=true, trim=0cm 0cm 0cm 0cm, width=.45\textwidth]{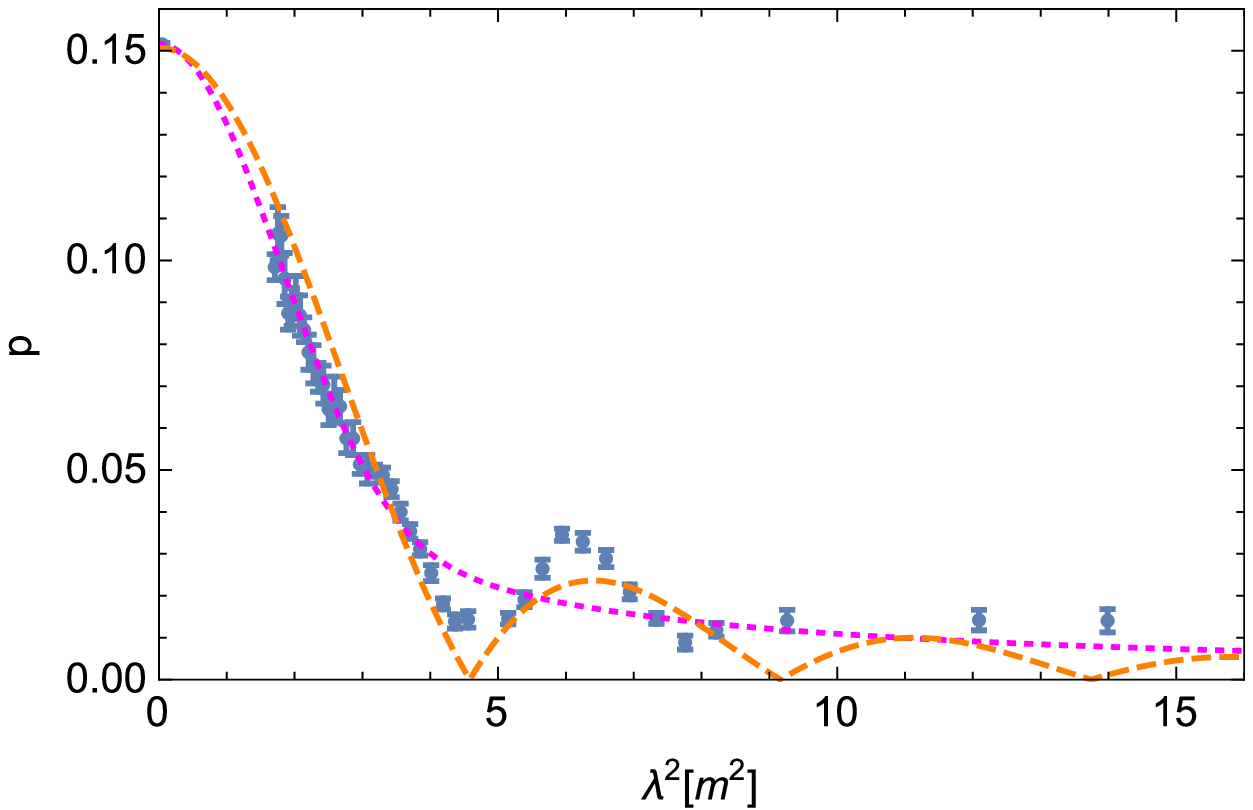}
  \caption{ Plot of the $p(\lambda^2)$ data for the southern hotspot (described in Figure~\ref{pwithleakage} and Section~\ref{sh}), 
  overlaid by the best-fitting one RM component models (Table~\ref{table:1RMcmpnt}). 
Top: Model (1a), red dashed line. Model (1b), green dotted line. Model (1c), blue solid line. 
Bottom: Model (1d), orange dashed line. Model (1e), magenta dotted line.}
\label{1RMmodels}
\end{figure}

Table~\ref{table:1RMcmpnt} lists the best-fitting one-RM-component model parameter values for the southern hotspot data, along with their reduced-chi-squared ($\chi^2_r$) and BIC values. 
Models (1a), (1b) and (1c) represent Eqn.~\ref{burneqn} for different combinations of the Faraday depolarization parameters ($\sigma_{\rm RM,B}$ and $\DeltaRM$, as can be inferred from the table entires themselves).
While Models (1d) and (1e) represent Eqn.~\ref{tribbleeqn}, with and without the $\DeltaRM$ parameter, respectively. The RM and intrinsic polarization angle are fixed using the values for `All' in Table~\ref{table:ATCAdata}, although variations in the parameters do not change the depolarization curve. 
As can be seen from the $\chi^2_r$ values and in Figure~\ref{1RMmodels}, where we overlay the best-fitting models on the  $p(\lambda^2)$ data, none of the one RM component models provide adequate descriptions of the data. Model~(1e) is selected by the BIC as the best-fitting model relative to the others. 
All the models generally describe the initial depolarization reasonably well, requiring a Faraday depolarization of $\DeltaRM \sim 0.7$\rad~or of $\sigmaRM \sim 0.2$\rad. However, most fail to adequately describe the secondary peak in $p(\lambda^2)$ at $\lambda^2 \sim 6$~m$^2$, as well as the data at longer wavelengths. 

\begin{table}
\centering
\caption{Best-fitting parameters for one RM component models.}
\label{table:1RMcmpnt}
\begin{tabular}{l c c c c c c}
\hline\hline
 & $p_0$ & $\sigma_{\rm RM,B}$ & $\DeltaRM$ & $\sigma_{\rm RM,T}$ & $\chi_r^2$ & BIC  \\
 & (\%) & (rad\,m$^{-2}$)     & (rad\,m$^{-2}$)  & (rad\,m$^{-2}$)  &                    &        \\ [0.5ex]
\hline
(1a)       & 15.1(2) & 0.233(7)          & --                   & --                      & 37         & 293   \\
(1b)       & 15.0(1) & --                     & 0.709(8)        & --                      & 16         & 258   \\
(1c)       & 15.1(1) & 0.05(1)            & 0.698(9)        & --                      & 14         & 255   \\
(1d)       & 15.0(1) & --                     & 0.69(1)          & 0.20(5)             & 14         & 258   \\
(1e)       & 15.2(1) & --                     & --                   & 0.27(1)             & 11         & 243   \\
\hline
\end{tabular}
\scriptsize{Note: ${\rm RM}=50$\rad~for all models.}
\end{table}

\begin{table*}
\centering
\caption{Best-fitting parameters for two RM component models. Overall best-fitting model in bold.}
\label{table:2RMcmpnts}
\begin{tabular}{l c c c c c c c c c c}
\hline\hline
Model     & $p_{0,1}$ & $\sigma_{\rm RM1,B}$ & $\Delta{\rm RM}_{1}$ & $\sigma_{\rm RM1,T}$ & $p_{0,2}$ & $\sigma_{\rm RM2,B}$ & $\Delta{\rm RM}_{2}$ & $\sigma_{\rm RM2,T}$ &$\chi_r^2$ & BIC  \\
              & (\%) & (rad\,m$^{-2}$)     & (rad\,m$^{-2}$)  & (rad\,m$^{-2}$)  & (\%) & (rad\,m$^{-2}$)     & (rad\,m$^{-2}$)  & (rad\,m$^{-2}$)  &                &          \\ [0.5ex]
\hline
(2a)       & 15.6(1) & 0.245(5)          & --                   & --              & 2.3(4)  & 0.05(2)             & --                   & --                     & 9.9         & 240.3   \\
(2b)       & 12.7(3) & --                     & 0.734(6)        & --              & 12.1(3) & --                     & 1.26(2)          & --                     & 3.2        & 190.8   \\
\bf{(2c)} & \bf{13.4(3)} & \bf{0.054(6)} & \bf{0.725(5)} & --        & \bf{11.4(4)} & --              & \bf{1.24(1)}          & --                & \bf{2.1}  & \bf{174.7}   \\
(2d)       & 11.2(8) & --                     & 0.72(1)          & --              & 13.4(6) & 0.237(5)          & --                   & --                     & 7.7         & 229.2   \\
(2e)       &  9.4(1) & --                      & 0.71(2)          & --              & 14.5(5) & --                     & --                   & 0.252(8)          & 7.0        & 227.9   \\
(2f)       & 12.7(3) & --                     & 0.726(7)       & 0.11(8)     & 12.2(4) & --                     & 1.24(2)          & --                     & 3.1        & 195.2   \\
\hline
\end{tabular}\\
\scriptsize{Note: $\psi_{0,1}=-96.0$\degr, $\psi_{0,2}=-43.6$\degr, and ${\rm RM_1}={\rm RM_2}=50$\rad~for all models.}
\end{table*}

The poor performance of the one RM component models, and the results from Section~\ref{atcadepol}, led us to try models with two RM components. 
As we are doubling the number of parameters, we expect the fits to improve significantly. However, the BIC (and $\chi^2_r$) strongly penalises additional free parameters, thus providing us with a robust statistical measure as to whether or not the two RM component models are truly better descriptions of the data than the one RM component models \citep[e.g.][]{schnitzeler2018}. 

\begin{figure} 
\centering
    \includegraphics[angle=0, clip=true, trim=0cm 0cm 0cm 0cm, width=.45\textwidth]{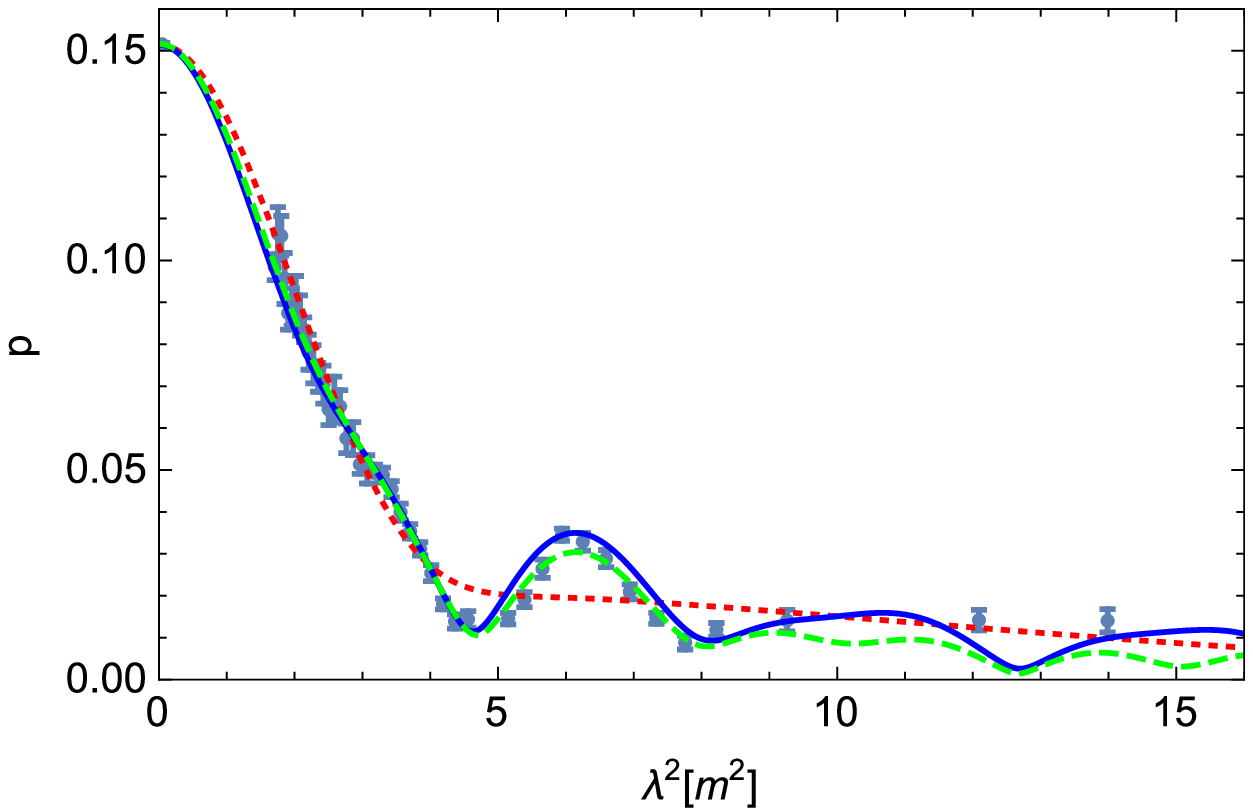}
    \includegraphics[angle=0, clip=true, trim=0cm 0cm 0cm 0cm, width=.45\textwidth]{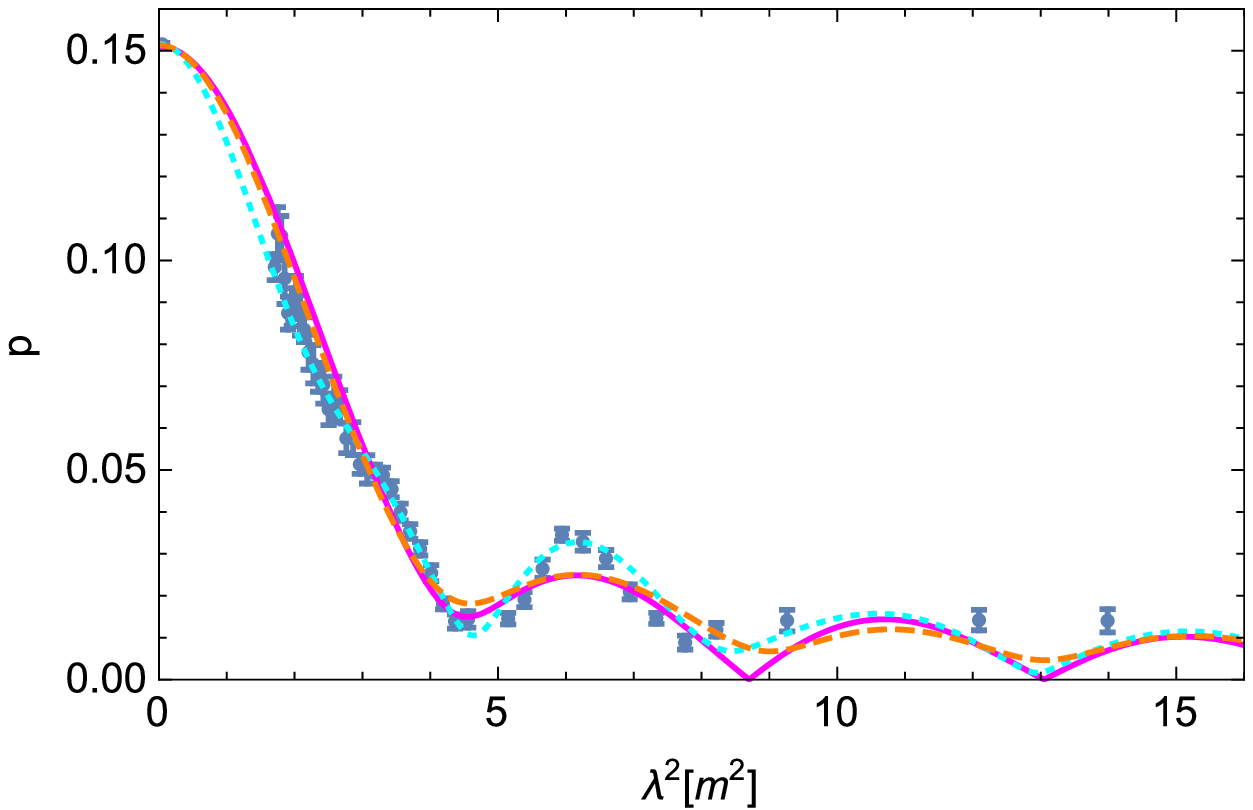}
  \caption{ Plot of the $p(\lambda^2)$ data for the southern hotspot (described in Figure~\ref{pwithleakage} and Section~\ref{sh}), 
  overlaid by the best-fitting two RM component models (Table~\ref{table:2RMcmpnts}). 
Top: Model (2a), red dotted line. Model (2b), blue solid line. Model (2c), green dashed line. 
Bottom: Model (2d), magenta solid line. Model (2e), orange dashed line. Model (2f), cyan dotted line. } \label{2RMmodels}
\end{figure}

Table~\ref{table:2RMcmpnts} lists the fitted parameter values of the two RM component models, and Figure~\ref{2RMmodels} shows the $p(\lambda^2)$ data overlaid by the best-fitting two RM component models. Models (2a), (2b), (2c) \& (2d) represent Eqn.~\ref{burneqn} for different combinations of the Faraday depolarization parameters ($\sigma_{\rm RM,B}$ and $\DeltaRM$), in the context of a two RM component model. While Models (2e) and (2f) represent Eqn.~\ref{tribbleeqn}, for different $\sigma_{\rm RM,T}$ and $\DeltaRM$ parameter combinations. 
Here we have used the information from Table~\ref{table:ATCAdata} to constrain the RM for both components to 50\rad~and the intrinsic polarization angles to $-96.0$\degr and $-43.6$\degr~(from Table~\ref{table:ATCAdata}). We also tried fits where ${\rm RM_1}$, $\psi_{0,1}$ and ${\rm RM_2}$, $\psi_{0,2}$ were allowed to vary around their respective values, but we did not find any significant improvements in the fits. 
 
Overall, Model~(2c) is the favoured best-fitting model with the lowest BIC and $\chi^2_r$ values, and provides a considerably better fit than the best-fitting one RM component model, Model~(1e). 
However, the value of $\chi^2_r=2.1$ indicates that Model~(2c) does not fully capture the data and/or the errors in the MWA $p(\lambda^2)$ data are underestimated (by a factor of $\sim$1.7). While a more in-depth error analysis of the MWA data may indeed increase the size of the error bars, the simplified assumptions in the depolarization models, as identified in Section~\ref{atcadepol}, are the most likely limiting factor for the model-fitting.  
We do not consider three RM component models (or more complex model geometries) mainly because of the large number of model parameters that would need to be fixed due to our inability to do full `qu-fitting' for the current dataset. 

Based on the results of Model~(2c), we tentatively identify Region E as the component with $\sigma_{\rm RM1,B} = 0.054\pm0.006$\rad~and $\Delta{\rm RM}_{1}= 0.725\pm0.005$\rad, and Region W with $\Delta{\rm RM}_{2}= 1.24\pm0.01$\rad. 
Reassuringly, the ratio of $p_{0,1}$ and $p_{0,2}$ is 0.85, which is very similar to the ratio of the $p_{\rm ATCA}$ values for Region E \& W in Table~\ref{table:ATCAdata}. The lower $p_{0,1,2}$ values are understandable in the context of the large synthesised beam of the MWA. 
In Section~\ref{interpret}, we use the results of Model (2c), in particular the Faraday depolarization parameters, to derive constraints on the properties of the magnetoionic material local to the hotspot. 


\subsection{Northern Hotspot and Bridge}
\label{nh}
Using the MWA and NVSS data, we find a total intensity spectral index of $\alpha=-0.78\pm0.02$ for the northern hotspot region, at 3\arcmin.3 resolution and centred on the peak polarized intensity of the northern hotspot. The average RM for the northern hotspot of $\sim$$+36$\rad~is significantly lower ($\sim$14\rad) than the southern hotspot. While it is possible that this RM difference is due to an asymmetry in the environment on either side of the host galaxy, we consider it more likely that the difference is due to a variation in the Milky Way RM on a scale of $\sim$15\arcmin. 
\footnote{The Galactic coordinates of the radio galaxy are ($l$, $b$) = (229.9$\degr$, -12.4$\degr$). } 

\begin{figure}
\begin{center}
\includegraphics[width=0.9\columnwidth]{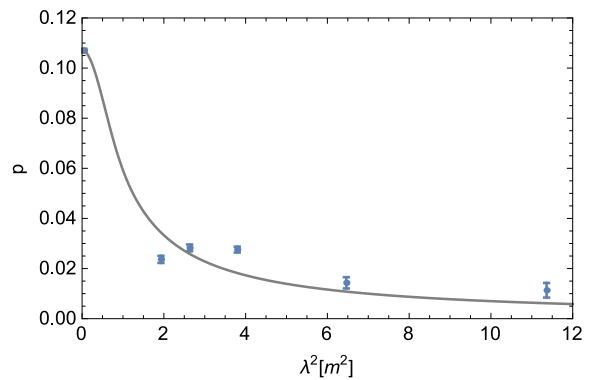} 
\caption{Plot of $p(\lambda^2)$ for the northern hotspot region (blue data points) from MWA and NVSS (see Section~\ref{nh}). 
The solid line shows a single RM component model with a Faraday depolarization of $\sigma_{\rm RM,T}\sim 0.9$\rad~and 
an intrinsic degree of polarization of $\sim$11\%.
}
\label{nhdepol}
\end{center}
\end{figure}

To analyse the depolarization behaviour, we use the MWA band-averaged polarization data (Table~\ref{table:rmsummary}), as the low signal to noise in polarization does not allow us to reliably extract sub-band information. Thus, we do not sample enough of the depolarization curve to accurately model the depolarization behaviour (Figure~\ref{nhdepol}). However, combining the MWA and NVSS data, and using the Tribble-law for depolarization (Eqn.~\ref{tribbleeqn}), we can roughly estimate the amount of Faraday depolarization as $\sigma_{\rm RM,T}\sim 0.9$\rad~ from an intrinsic degree of polarization of $\sim$11\% (Figure~\ref{nhdepol}). 
We do not have enough data points to reliably constrain a two RM component model. 
More data is required between 1.4 GHz and 300 MHz, along with more sensitive low frequency data, to constrain the Faraday structure of the northern hotspot in a similar manner as the southern hotspot. 

We also detect faint polarized emission from the northern bridge, at J2000 06$^{\rmn{h}}$36$^{\rmn{m}}$32$\fs$8, $-20$\degr 32$\arcmin$15$\arcsec$. The polarized emission (found at $\sim$+38\rad) is fainter than the sidelobes from the southern hotspot at this location. However, the peaks are sufficiently separate in Faraday depth space to estimate a degree of polarization of $\sim$1.2\% at 216 MHz and $\sim$0.6\% at 185 MHz (it is undetected in the lower frequency bands). We attempted to improve the quality of the FDF at this location by deconvolving the $I$, $Q$, $U$ images averaged over 0.64 MHz. However, this did not improve the FDF at this location in a significant manner because for these observations the MWA beam is not well constrained at the location of the source. From the NVSS $I$, $Q$, $U$ data smoothed to the MWA resolution, we find this region is $\sim$10\% polarized at 1.4~GHz. A Faraday depolarization of $\sigma_{\rm RM,T}\sim 1.0$\rad~can decrease this to the observed degree of polarization in the MWA band.

\begin{figure}
\begin{center}
\includegraphics[clip=true, trim=0cm 2cm 0cm 2.5cm, width=0.95\columnwidth]{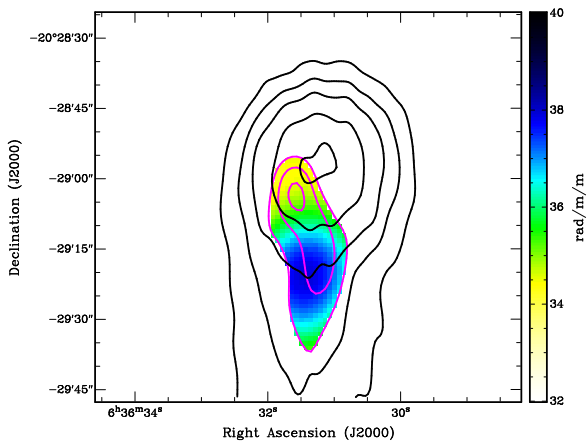} 
\caption{Total intensity (black contours), polarized intensity (magenta contours) and rotation measure distribution (colour-scale) of the northern hotspot at 25\arcsec~resolution with the ATCA at 16~cm. 
}
\label{RMn}
\end{center}
\end{figure}

The ATCA data (Figure~\ref{RMn}) show that bright polarized emission is present south-east of the peak in the total intensity emission. The polarized emission extends further south along the lobe structure until it is too faint to reliably detect. The RM varies across this region from approximately $+34$\rad~to $+38$\rad~(with a degree of polarization varying from $\sim$16\% to $\sim$37\% in a north-south direction). 
Observations with better $uv$-coverage and sensitivity are required to study the resolved polarization and RM structure of the northern hotspot and lobe in more detail. 

\section{Physical interpretation of the observed Faraday depolarization}
\label{interpret}
In determining the magnetoionic properties of the gas local to the radio galaxy, we focus solely on the results from the southern hotspot. 
We consider the two main possible scenarios to explain the Faraday depolarization results, in the context of the expected hotspot environment (Figure~\ref{fig:cartoon}): internal Faraday rotation (from low-energy relativistic electrons, or from thermal gas internal to the synchrotron emitting region), and external Faraday rotation from magnetised thermal gas outside the radio galaxy. 

\subsection{Internal Faraday depolarization}
Ideally X-ray data would be used to constrain the magnetic field strength in the hotspot without the need for the assumption of equipartition \citep{harris1994}, but X-ray data for the southern hotspot are currently not available. However, several studies of hotspots in high-luminosity FRIIs find that synchrotron self-compton (SSC) models can explain the radio and X-ray hotspot data, with hotspot magnetic field strengths similar to those derived from the equipartition/minimum energy condition \citep{hardcastle1998b, hardcastle2002}. In low-luminosity FRIIs the comparison is more complicated due to additional X-ray synchrotron emission \citep{hardcastle2004, hardcastle2016}.

We use the synchrotron minimum energy (equipartition) magnetic field formulation described in \cite{worrallbirkinshaw2006}, which recasts the traditional formulation based on the maximum and minimum radio frequencies \citep[e.g.][]{longair2011} to the more physically meaningful low and high relativistic-electron energy distribution cutoffs ($\gamma_{\rm min}$, $\gamma_{\rm max}$).  
We find an equipartition magnetic field strength of $B_{\rm eq}\sim20$\,$\mu$G, using the observed radio flux for the southern hotspot (Table~\ref{table:rmsummary}) and the best-fitting spectral index of $\alpha=-0.8$, considering the hotspot as a sphere of radius 10~kpc. 
Based on SSC model results for hotspots \citep{hardcastle2004, hardcastle2016}, we assume a hotspot volume completely filled with relativistic leptons (filling factor $f=1$ and $k=0$)\footnote{$k$ is the ratio of total energy in non-radiating particles to that in synchrotron-emitting leptons.} with $\gamma_{\rm min}=1000$. 

We cannot constrain the value of $\gamma_{\rm min}$ from our observations as we do not detect any evidence of a turnover of the total intensity spectrum of the hotspot at the lowest frequencies (Fig.~\ref{SHspix}), which could indicate the presence of a low-energy cutoff in the relativistic electron energy distribution \citep[e.g.][]{carilli1991}. However, many studies suggest that $\gamma_{\rm min}$ is somewhere between 100 and 1000 for FRII hotspots \citep{carilli1991, hardcastle1998a, godfrey2009, mckean2016}.  
Moderate filling factors and a proton population that does not dominate the energetics are still consistent with SSC model predictions  \citep{hardcastle2004}, so in principle a magnetic field strength of up to $B_{\rm eq}\sim70$\,$\mu$G may still be possible ($f=0.1$, $k=1$, $\gamma_{\rm min}=100$). However, based on the most likely parameters from the SSC modelling and for simplicity, we use $B_{\rm eq}=20$\,$\mu$G. 

\subsubsection{Internal Faraday depolarization from thermal gas}
\label{backflow}
Ordered polarization structures in FRII hotspots (and elsewhere in radio galaxies) are most commonly considered to arise from the stretching and compression of an initially disordered magnetic field by the flow of the relativistic plasma \citep{laing1980}. 
Although current three dimensional MHD simulations indicate that large-scale ordered fields likely exist in the jet launching region \citep[e.g.][]{mckinneyblandford2009} and possibly persist to kiloparsec scales \citep{sashabromberg2016}, there is of yet no strong observational support for this scenario in relation to radio galaxy hotspots. Therefore, we assume the partially ordered field responsible for the observed degree of polarization is effectively random in the context of the line of sight magnetic field ($B_{||}$) responsible for the Faraday rotation. 

Internal Faraday rotation may occur within the hotspot itself or in the surrounding lobe material. Based on the assumptions used to estimate $B_{\rm eq}$, it is unlikely that the hotspot contains a significant amount of thermal plasma. However, significant amounts of thermal plasma could in principle exist in the surrounding lobe material. As the supersonic jet terminates at the hotspot region, the shock-accelerated material flows backwards to inflate an over-pressured cocoon/lobe surrounding the jet (Figure~\ref{fig:cartoon}). If this `backflow' entrains thermal gas from the environment, through surface-wave instabilities for example \citep{bicknell1990}, then it may be a significant source of Faraday depolarization. 

For $\Delta{\rm RM} \sim 0.7$\rad~due to internal Faraday rotation, then following Eqn.~\ref{rmeqn}, we estimate an average internal thermal electron density of $n_e \sim 10^{-4}\, (N_{\rm rev}/1000)^{1/2}\,\,{\rm cm}^{-3}$, 
using a path length of twice the radius of the hotspot, $B_{||}=B_{\rm eq}/\sqrt 3$ and assuming $B_{||}$ reverses direction of order 1000 times along the path length ($N_{\rm rev}$). 

\subsubsection{Internal Faraday depolarization from low-energy relativistic electrons}
In light of the small observed Faraday depolarization, we now consider internal Faraday rotation from purely relativistic particles in the hotspot. In this regard, it is the relativistic electrons at the low energy end of the electron energy spectrum that contribute most to the internal Faraday rotation. 
As there is no Faraday rotation in a pure pair plasma, we require an excess of low-energy electrons (or positrons). 
Following \cite{jonesodell1977} and \cite{homan2012}, we have the RM from relativistic particles (RM$_r$), 
\begin{equation}
{\rm RM}_r \sim \frac{ 1.6 \log{\gamma_{\rm min}} }{ \gamma_{\rm min}^2 } \int_L^0 n_r B_{||} \,dl  
\label{rrmeqn}
\end{equation}
for $\alpha=-0.8$ and where $n_r$ is the relativistic particle number density and all units are the same as before. 
Using the equipartition condition, we can write the relativistic particle density in terms of $\gamma_{\rm min}$ 
and the magnetic field strength, giving $n_r\sim 1.8\times10^{11} B^2 \gamma_{\rm min}^{-1}$~cm$^{-3}$ \citep[e.g.][]{godfrey2012}.  
For $\gamma_{\rm min}=1000$ and $B_{\rm eq}=20$~$\mu$G we find a negligible value of RM$_r\sim10^{-8}$\rad.
Thus, this scenario can be ruled out for self consistent parameters related to the hotspot emission.

\subsection{External Faraday depolarization}
The most common explanation for Faraday depolarization in radio galaxies is depolarization caused by turbulent, magnetised thermal gas in the intragroup/intracluster medium in which the radio galaxy is expanding into \citep[e.g.][]{laing2008}. 
In the case of the southern hotspot, we have a compact emission region at a distance of $\sim$500~kpc from its host galaxy, in an isolated environment \citep{Baum:1988}. 

Without deep X-ray observations to reliably determine the external particle number density, we are left to estimate it by other means. 
A plausible particle number density for the intergalactic medium at the distance ($r$) of the hotspot, can be estimated using a standard profile $n_{\rm IGM}(r) \sim n_0(r/a_0)^{-b}$. This gives $n_{\rm IGM} (500\,\,{\rm kpc}) \sim 3\times10^{-5}$~cm$^{-3}$, for the typically derived values of $n_0 \sim 10^{-2}$~cm$^{-3}$, $a_0 \sim 10$~kpc, and $b \sim 1.5$ \citep{mulchaey1998, sun2012}. However, the uncertainty in the applicability of these scaling-relations to such an isolated galaxy system as PKS~J0636-2036, and the large scatter in the scaling parameters, means that this external gas number density value is quite crude, and possibly overestimated. In the case that this radio galaxy extends into an extremely low density environment, then it may be propagating in the warm-hot intergalactic medium (WHIM), which has an expected particle density of $n_{\rm WHIM} \sim 10^{-6}$~cm$^{-3}$ \citep[e.g.][]{subrahmanyan2008, machalski2011}. 

\subsubsection{External Faraday depolarization due to a magnetised IGM}
In order to estimate physical quantities from the $\DeltaRM$ parameter, we need to include the correction factor of 1.35 for the case of a linear RM gradient in a Gaussian beam profile (Section~\ref{polmodel}). For Model~(2c), this gives $\Delta{\rm RM}_{1}^c \sim 0.5$\rad~and $\Delta{\rm RM}_{2}^c \sim 0.9$\rad~(where the $c$ superscript denotes the `corrected' values). 
Explaining $\DeltaRM \sim 0.5$ to $0.9$\rad~requires uniform field structures on scales of at least 10~kpc, to produce a smooth gradient in the RM across the emission region. A cell size of 10~kpc and a path length of 500~kpc requires an IGM magnetic field strength ($B_{\rm IGM}$) of $\sim$$0.3$ to $0.5\,\mu$G for $n_{e} \sim 3\times10^{-5}$~cm$^{-3}$. If the outer scale of the field extends to 500~kpc then $B_{\rm IGM}\sim0.04$ to $0.07\,\mu$G.
If $n_{\rm WHIM} \sim 10^{-6}$~cm$^{-3}$ is more appropriate for the IGM at the distance of the hotspot, then the required magnetic field strengths are $\sim$30 times larger. 

\subsubsection{External Faraday depolarization due to shock-enhanced IGM gas}
\begin{figure}
\centering
\includegraphics[width=\linewidth]{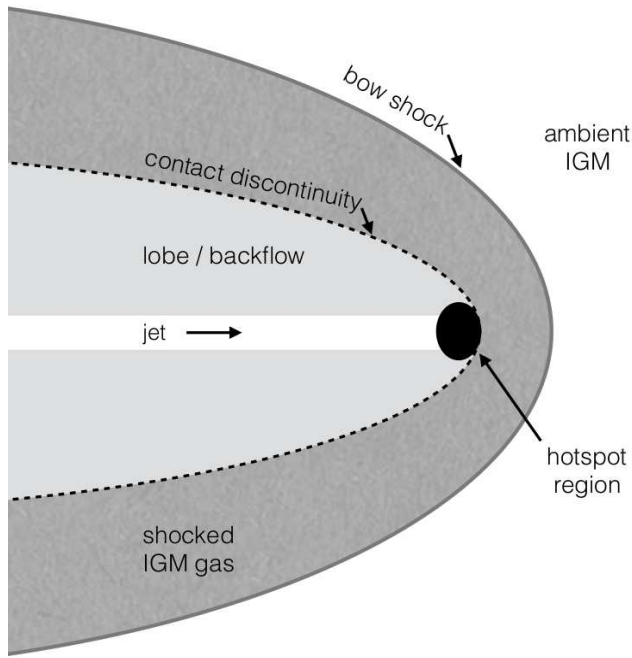}
\caption{Cartoon illustrating the various components near the termination point of an FRII radio galaxy, as described in Sections~\ref{interpret}, to interpret the observed Faraday depolarization.}
\label{fig:cartoon}
\end{figure}

An alternative scenario for external Faraday depolarization is from ambient gas that has been compressed and heated by the bow shock from the advancing hotspot (see Figure~\ref{fig:cartoon}). While strong bow shocks around FRII radio galaxies are expected due to the supersonic advance of the hotspot \citep{kaiseralexander1997}, there is of yet little evidence for such strong shocks surrounding powerful radio galaxies (although see  \cite{stawarz2014} for a scenario for why these shocks might be difficult to detect in X-ray observations). 

Evidence for this scenario has been presented previously in the case of the FRII radio galaxy, Cygnus A \citep{carilli1988}. They associated an approximately `hemispherical structure of large RM' with the hotspot in the eastern lobe of Cygnus A, and considered this as being due to the compression of the IGM by the expected bow shock. 
Furthermore, as studied by \cite{guidetti2011,guidetti2012}, compression alone of ambient gas is insufficient to explain the systematic RM gradients (or `RM bands') seen in their observations of FRI radio galaxies. They find that `magnetic draping' \citep{dursipfrommer2008}, where the external magnetic field is swept up and aligned by the expanding radio source, is required (in addition to some compression). 

In our case, considering a maximum density compression factor of 4, gives an enhanced external gas number density of $n_e\sim 1.2\times10^{-4}$~cm$^{-3}$. 
The extent of the bow-shocked region is unclear but rough estimates can be made based on numerical simulations of jets in which an extent of approximately 1 to 2 times the size of the hotspot is found \citep{norman1982}. Assuming that the region of enhanced density and magnetic field extends $\sim20$~kpc outside the edge of the radio galaxy, and if the line-of-sight magnetic field does not reverse direction, then we find a shock-enhanced $B_{||}$ ranging from $\sim$0.3 to 0.5~$\mu$G (for $\DeltaRM \sim 0.5$ to $0.9$\rad). 

This scenario can be quite attractive if the advancing bow-shock has significantly enhanced the gas density and magnetic field strength in this region, in addition to the magnetic-draping effect that generates the initial uniform magnetic field. 
However, if the IGM density is similar to that of the WHIM ($\sim10^{-6}$~cm$^{-3}$), then the required uniform line-of-sight magnetic field strength in the bow-shocked region becomes quite large ($\sim10$~$\mu$G).

\section{Discussion}
\label{sec:discussion}

\subsection{Internal or external Faraday depolarization}
\label{depol_discussion}
For internal Faraday rotation from thermal gas, we estimated a thermal electron density of $n_e\sim 10^{-4}$~cm$^{-3}$ in the backflow/lobe region surrounding the hotspot (Section~\ref{backflow}). The expected temperature of thermal gas internal to an FRII hotspot/lobe region is unknown. From observations of X-ray cavities associated with radio galaxies, we expect $T > 10$~keV \citep{schmidt2002, gitti2007} and $T > 500$~keV from models of FRI radio galaxies \citep{crostonhardcastle2014}. However, these observations and models do not directly apply to FRII radio galaxies whose jets are not expected to entrain and heat much, if any, thermal material. 
In any case, if we assume a gas temperature of $T=500$~keV, then the internal pressure from thermal material is $p_{\rm th} \sim 2n_e k_B T \sim 3\times10^{-11}$~Pa. We can compare this value to the synchrotron gas pressure of $p_{\rm min} \sim 10^{-12}$~Pa, which is $\sim$3 times smaller than $p_{\rm th}$. 
This is inconsistent with the expectation of thermal gas not being energetically significant in FRII lobes \citep{croston2005, ineson2017}, as opposed to FRI radio galaxies where the opposite is expected to be true \citep{croston2008, birzan2008, crostonhardcastle2014}. Therefore, we consider the case of significant internal Faraday rotation from thermal gas as unlikely.

For the case of external Faraday depolarization, we lack direct tracers of the ambient thermal gas density. If our extrapolation from group environment scaling relations is close to accurate, then IGM magnetic field strengths ranging from 0.04 to 0.5\,$\mu$G, for outer and inner field scales of 500~kpc and 10~kpc, respectively, can explain the observed amounts of Faraday depolarization. 
In both cases, the estimated field strengths would increase by a factor of $\sim$30 if the external gas number density is similar to that expected for the WHIM. Such high field strengths would be inconsistent with recent upper limits for intergalactic filaments \citep{brown2017, vernstrom2017}.
Therefore, an ambient gas density of order $10^{-5}$~cm$^{-3}$ is probably more realistic than a WHIM density of $\sim10^{-6}$~cm$^{-3}$. 
A scenario in which an advancing bow-shock has significantly enhanced the gas density and magnetic field strength over a short range in the ambient medium can also explain the observations, with field strengths in the range 0.3 to 0.5\,$\mu$G. 

From this analysis we can conclude that external Faraday rotation is most likely the dominant factor in the observed depolarization, although the external field strength and scales are not very well constrained. Improved constraints may be obtained from high resolution radio observations at low frequencies, which would better map the RM variations and depolarization across the hotspot region (instead of relying on the depolarization modelling of the integrated hotspot emission). 
Deep X-ray observations could help to better determine the magnetic field strength in the hotspot, however detecting the ambient thermal gas (shocked or not) would be extremely challenging for current X-ray telescopes. 

\subsection{A giant radio galaxy}
The observed linear size of PKS~J0636$-$2036 on the sky is $\sim$957~kpc, putting it just short of the usual size of at least 1~Mpc for classification as a Giant Radio Galaxy (GRG). 
The axial ratio ($R_T$, lobe length divided by half the lobe width) of the source ranges from $\sim$5 (nearest the host galaxy) to $\sim$10 (near the outermost regions). This is more similar to that measured for double-double radio galaxies (DDRGs) with $9\lesssim R_T\lesssim16$ than ``normal'' FRIIs with $2\lesssim R_T\lesssim5$ \citep[][]{machalski2011}. DDRGs are indicative of recurrent jet activity in which the restarted jet propagates in an underdense environment leading to their large axial ratios. 
In applying this to GRGs, \cite{machalski2011} raised the possibility of diffuse emission beyond the hotspot of GRGs with large axial ratios. This would mean that the implied low density of the IGM surrounding GRGs was as a result of a previous epoch of jet activity. 
However, while not directly invalidating this model, we do not see any evidence for diffuse emission beyond the hotspots, for a noise level of $\sigma\sim0.15$~Jy~beam$^{-1}$. 
Alternatively, the large axial ratio may be due to unusual jet stability \citep{scheuer1974, scheuer1982} enabling the jet to advance faster and further into the surrounding medium. Indeed the intrinsically poor environment (in terms of the expected ambient gas density of the IGM) may substantially assist in achieving both the unusual jet stability and large linear size. 

The southern lobe is significantly longer than the northern lobe, with a size ratio of $\sim$1.3 between the southern and northern lobe. Such an asymmetry in length can be explained by light travel time effects in dynamical models of FRII sources \citep[e.g.][]{longair1979}. The expected size ratio $q=(1+\beta \cos \theta)/(1-\beta \cos \theta)$, where $\beta=v_h/c$ is the advance speed ($v_h$) as a fraction of the speed of light ($c$), and $\theta$ is the angle to the line of sight. Using the measured value of $q=1.3$, then $\theta$ must be less than $82.5\degr$. For typical advance speeds of $0.15c$ to $0.3c$ \citep{best1995}, then $\theta$ ranges from $30\degr$ to $65\degr$ giving de-projected total linear sizes ranging from 1.87~Mpc to 1.03~Mpc, respectively. 

However, it may not be the case that the light travel time is the only effect in contributing to the observed size ratio. 
From a large study of the emission line properties of powerful radio galaxies, \cite{mccarthy1991} found that the more intense emission line gas was on the shorter arm side of the radio galaxy. This indicates that the radio galaxy environment has a potentially important contribution to any asymmetries in the relative lengths of the lobes. The host galaxy of PKS~J0636$-$2036 has extended emission line gas within 20\arcsec~of the host nucleus with an unclear relation to the jet structure \citep{Baum:1988}, although a filament to the north may be evidence of a previous interaction between the northern jet and its environment, potentially limiting the growth of the northern lobe compared to the southern lobe. The northern lobe also has a relatively bright emission feature about 240\arcsec~from the core, indicative of significant dissipation of the jet energy in this region, which might also have limited the advance of the northern lobe into the IGM (as compared to the southern jet which appears to dissipate the majority of its energy at the southern hotspot). 

In any case, whether or not the true linear size of the radio galaxy is greater than 1~Mpc, it almost certainly will be in the future, as the bright, compact hotspots indicate that this radio galaxy is still  expanding (supersonically) into the IGM. Giant radio galaxies that are observed in the relict phase with a more relaxed, diffuse structure are more likely to be in pressure equilibrium with the IGM and near the end of their life \citep[e.g.][]{subrahmanyan2008}. 

\subsection{Finding more polarized radio galaxies at low frequencies}
The current number density of extragalactic polarized sources with the MWA is quite small, estimated at $\sim$1/100~deg$^{-2}$ in \cite{Lenc:2016, lenc2017}. 
Here we briefly comment on some of the reasons why PKS~J0636$-$2036 is detected in polarization across the full MWA 
frequency coverage (70 to 230~MHz), along with the future prospects for detecting more polarized source at low frequencies. 
 
Firstly, the large angular size of the source ($\sim$15\arcmin) means that it is well resolved, given the MWA beamsize of $\sim$3.5\arcmin, even though significant structure remains on smaller scales within the hotspot (as identified by the 5\arcsec~ATCA data).  
Therefore, the effect of beam depolarization is reduced, with regions of different intrinsic polarization and/or Faraday rotation 
properties being separated on the sky. 

Secondly, and likely more importantly, the intrinsically large linear size of the source ($\sim$957~kpc) means that it extends well outside the host galaxy environment and into the tenuous intergalactic medium. This minimises the impact of external Faraday dispersion as a means 
of strongly depolarizing the emission, as observed in group and cluster environments \citep[e.g.][]{laing2008}. 
Additionally, \cite{strom1973} found that sources with large angular and linear size had lower depolarization and found that this was most likely due to Faraday dispersion caused by the hot gas environment surrounding the host galaxy \citep{stromjagers1988}. 
As discussed in \cite{farnsworth2011}, even small Faraday dispersions of $\gtrsim$\,1\rad~are sufficient to depolarize emission below the typical observational detection limits at low frequencies.

Furthermore, the FRII morphology of PKS~J0636$-$2036 means that the brightest features are at the extremities of the radio source 
(i.e.~in the least dense region of the environment), and they are also compact (i.e.~the variation in RM across the emission 
region will be relatively small). This further helps to minimise the effect of external Faraday dispersion. 
In support of this, a recent catalog of $\sim$90 polarized sources at 150~MHz by van Eck et al.~(in preparation), finds that the majority of sources are associated with hotspots of FRIIs. 

Thirdly, the large Galactic RM of this source ($\sim$35 to 50\rad) means that the polarized emission in the Faraday dispersion function 
is shifted away from the instrumental polarization near ${\rm RM}\sim0$\rad, making it easier to reliably identify the real emission 
from the source; although \cite{Lenc:2016} circumvented this problem somewhat by using the time-variable RM of the ionosphere. 

This means that the most likely candidates for the detection of extragalactic polarized emission at low frequencies are FRII morphology 
radio galaxies with large linear sizes that are in low galaxy density environments. Low frequency telescopes with high angular resolution 
are likely to find many more sources than currently possible with the MWA, given the extra constraint for the MWA of large angular size sources. 
The high angular resolution currently provided by LOFAR of $\sim$6\arcsec~\citep{shimwell2017} indicates that a much greater number density of extragalactic 
polarized sources can be detected, possibly up to one polarized sources every three square degrees (\cite{Mulcahy:2014}, Neld et al.~in prep). 
In the future, the low-frequency component of the SKA can provide further improvements on the ability to detect polarized sources at low frequencies due to improved survey sensitivity and minimal radio frequency interference. However, we expect the ability to obtain high angular resolution images, similar to LOFAR, will be a key additional requirement. 

\section{Conclusions}
\label{sec:conclusion}

We have presented a broadband polarization and Faraday rotation study of the nearby radio galaxy PKS~J0636$-$2036 ($z=0.0551$), 
using data from the MWA at 3\arcmin.3 resolution from 70 to 230~MHz and the ATCA at 5\arcsec~resolution from 1 to 3 GHz.  
This FRII radio galaxy has a linear size of $\sim$957~kpc, extending well outside its isolated ellliptical 
host galaxy and into the tenuous intergalactic medium. We find no evidence for flattening or a turnover in the total intensity spectrum of 
either hotspot, down to $\sim$75~MHz. 
A total intensity spectral index of $\alpha \sim -0.78$ was found for the northern hotspot, while for the southern hotspot 
$\alpha \sim -0.82$. 

We have detected polarized emission from both the northern and southern hotspot of this radio galaxy, as well as part of the northern bridge. 
The northern hotspot is $\sim$11\% polarized at short wavelengths (with the 
NVSS at 1.4~GHz), and decreases from $\sim$3\% to 1\% across the MWA band, with an RM of $\sim+$35\rad~and an RM dispersion of $\sim$0.9\rad. 
The polarized emission from the northern bridge is detected down to 185~MHz, and is consistent with a Faraday dispersion of  $\sim$1\rad.
The degree of polarization of the southern hotspot varies from $\sim$9\% to $\sim$1\% across the MWA band, at an RM of $\sim$50\rad, and it is $\sim$15\% polarized in the NVSS. The depolarization is broadly consistent with an RM dispersion of $\sim$0.7\rad. 

This is the first time the broadband polarization behaviour of a radio galaxy has been determined at such low frequencies. 
The extensive wavelength-squared coverage provided by these observations ($\sim$1.7 to 16~m$^2$) demonstrates the increased precision with which Faraday rotation properties can be determined, compared to traditional cm-wavelength facilities (e.g.~VLA, ATCA). 
However, even though the hotspots are resolved from the lobes by the MWA, substantial structure remains within the southern hotspot. The 5\arcsec~imaging with the ATCA reveals two main sub-regions with high degrees of polarization ($\lesssim45\%$) and an intrinsic magnetic field structure aligned with the edges of the hotspot region. 

A general purpose, polarization model-fitting procedure was applied to the data from the southern hotspot to determine the Faraday 
depolarization parameters.
We find that single RM component models cannot describe the broadband depolarization data from this region. 
Two RM component models provide substantially better fits to the data, although the best-fitting model still does not fully describe the observed data ($\chi^2_r=2.1$). This is likely due to simplifying assumptions in the input models. In particular, the assumption that each RM component is well characterised by a constant intrinsic polarization angle. This is inconsistent with the structure revealed by the ATCA.
This illustrates the need for high angular resolution at low frequencies as well as the development and application of more detailed polarization models.

We find that a magnetised IGM is more likely to be responsible for the majority of the observed depolarization than Faraday rotation internal to the hotspot emission region. However, due to the poorly constrained external gas density, we cannot uniquely determine the origin of the 
observed Faraday depolarization. For an estimated ambient density of $3\times10^{-5}$~cm$^{-3}$, IGM magnetic field strengths ranging from 0.04 to 0.5\,$\mu$G are consistent with the observed depolarization, depending on the exact scale size of the fluctuations and the location of the dominant Faraday rotation region (e.g.~in a compressed region outside the hotspot or distributed more uniformly in the IGM). 

Overall, this work shows the importance of low frequency polarization and Faraday rotation observations in the study of radio galaxies, the impact on their environments and the magnetised properties of the intergalactic medium. 
Current and future low frequency surveys with LOFAR, MWA-Phase 2 and SKA1-low, should find a much greater number density of extragalactic 
polarized sources. This will enable powerful statistical studies of the physical properties of the radio galaxy population and the magnetisation 
of the intergalactic medium.

\section*{Acknowledgements}
S.P.O acknowledges financial support from UNAM through the PAPIIT project IA103416 and from the Deutsche Forschungsgemeinschaft (DFG) under grant BR2026/23. 
This research was conducted by the Australian Research Council Centre of Excellence for All-sky Astrophysics (CAASTRO), through project number CE110001020. 
This scientific work makes use of the Murchison Radio-astronomy Observatory, operated by CSIRO. We acknowledge the Wajarri Yamatji people as the traditional owners of the Observatory site. Support for the operation of the MWA is provided by the Australian Government (NCRIS), under a contract to Curtin University administered by Astronomy Australia Limited. We acknowledge the Pawsey Supercomputing Centre which is supported by the Western Australian and Australian Governments. The Dunlap Institute is funded through an endowment established by the David Dunlap family and the University of Toronto. We are grateful to \L{}ukasz Stawarz for helpful discussions and to the referee, Paddy Leahy, for helpful comments on the original version of this paper.

\bibliographystyle{mnras}
\bibliography{bflux_bib} 

\bsp	
\label{lastpage}
\end{document}